\documentclass[10pt,conference]{IEEEtran}
\usepackage{cite}
\usepackage{amsmath,amssymb,amsfonts}
\usepackage{algorithmic}
\usepackage{graphicx}
\usepackage{textcomp}
\usepackage{xcolor}
\usepackage[hyphens]{url}
\usepackage{fancyhdr}
\usepackage{hyperref}

\usepackage[section]{placeins}

\usepackage{subfig}
\usepackage{siunitx}
\usepackage{multirow}
\usepackage{booktabs}
\usepackage[normalem]{ulem} % keeps \em and \emph italic
\usepackage{xspace}

\newcommand{\blueHL}{\textcolor{black}}
\newcommand{\arch}{HETRI\xspace}

\usepackage{algorithm}

\newcommand{\hpcayear}{2027}

\newcommand{\hpcasubmissionnumber}{3216}
\title{HETRI: Heterogeneous Ising Multiprocessing}
\def\hpcacameraready{}

\newcommand\hpcaauthors{
Ahmet Efe,
H\"{u}srev C{\i}lasun,
Abhimanyu Kumar,
Ziqing Zeng,
Nafisa S. Prova,\\
Chris H. Kim,
Sachin S. Sapatnekar,
and Ulya R. Karpuzcu
}

\newcommand\hpcaaffiliation{
University of Minnesota, Twin Cities
}

\newcommand\hpcaemail{
\{efe00002, cilas001, kumar663, zeng0083, prova026, chriskim, sachin, ukarpuzc\}@umn.edu
}

\author{
  \ifdefined\hpcacameraready
    \IEEEauthorblockN{\hpcaauthors{}}
      \IEEEauthorblockA{
        \hpcaaffiliation{} \\
        \hpcaemail{}
      }
  \else
    \IEEEauthorblockN{\normalsize{HPCA \hpcayear{} Submission
      \textbf{\#\hpcasubmissionnumber{}}} \\
      \IEEEauthorblockA{
        Confidential Draft \\
        Do NOT Distribute!!
      }
    }
  \fi 
}

\fancypagestyle{camerareadyfirstpage}{%
  \fancyhead{}
  
  \fancyhead[C]{
    \ifdefined\aeopen
    \parbox[][12mm][t]{13.5cm}{\hpcayear{} IEEE International Symposium on High-Performance Computer Architecture (HPCA)}    
    \else
      \ifdefined\aereviewed
      \parbox[][12mm][t]{13.5cm}{\hpcayear{} IEEE International Symposium on High-Performance Computer Architecture (HPCA)}
      \else
      \ifdefined\aereproduced
      \parbox[][12mm][t]{13.5cm}{\hpcayear{} IEEE International Symposium on High-Performance Computer Architecture (HPCA)}
      \else
      \parbox[][0mm][t]{13.5cm}{\hpcayear{} IEEE International Symposium on High-Performance Computer Architecture (HPCA)}
    \fi 
    \fi 
    \fi 
    \ifdefined\aeopen 
      \includegraphics[width=12mm,height=12mm]{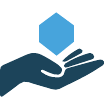}
    \fi 
    \ifdefined\aereviewed
      \includegraphics[width=12mm,height=12mm]{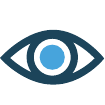}
    \fi 
    \ifdefined\aereproduced
      \includegraphics[width=12mm,height=12mm]{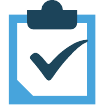}
    \fi
  }
  \fancyfoot[C]{}
}
\begin{document}
\maketitle

%Enables the camera ready header and footer
\ifdefined\hpcacameraready 
  \thispagestyle{camerareadyfirstpage}
  \pagestyle{empty}
\else
  \thispagestyle{plain}
  \pagestyle{plain}
\fi

\newcommand{\hpcaheight}{0mm}
\ifdefined\eaopen
\renewcommand{\hpcaheight}{12mm}
\fi

% arXiv version: remove HPCA running header
\thispagestyle{plain}
\pagestyle{plain}

%%%%%%%%%%%%%%%%%%%%%%%%%%%%%%%%%%%%%%%%
%%%%%%%% -- PAPER CONTENT STARTS -- %%%%
%%%%%%%%%%%%%%%%%%%%%%%%%%%%%%%%%%%%%%%%

\begin{abstract}
Ising machines are effective solvers for combinatorial optimization
problems, mapping optimal solutions to the minimum energy states of a
physical system -- a network of discrete spins that naturally converges
to equilibrium upon perturbation. The performance of an Ising machine
is dictated by how well the machine capacity matches the problem size,
and how well the machine connectivity can express the problem variable
interactions. Densely connected machines can directly support arbitrary
variable interactions but incorporate far fewer physical spins under
the same hardware budget; sparsely connected machines accommodate more
spins but require extra physical spins to bridge connectivity
mismatches, rendering a larger and oftentimes harder problem to solve.

\blueHL{To minimize mismatch between problem and machine
characteristics, this paper makes the case for \arch: Heterogeneous
Ising Multiprocessing, which integrates independent Ising cores of
diverse connectivity on the same chip and routes each problem to the
core whose connectivity best matches its structure. Which cores such a
chip should carry cannot be settled by reasoning alone, so we build the
flow that answers it: a digital Ising emulator that fixes the spin and
coupler design while varying only connectivity, a solver study that
establishes what can run at workload scale, and scheduler approaches
that perform the problem--core matching. We evaluate on NP-complete and
NP-hard benchmarks spanning a wide range of problem
connectivity. Under the very same hardware
budget and technology, the best \arch\ configuration outperforms the
homogeneous alternatives on every workload tested.}
\end{abstract}

\section{Introduction} \label{sec:intro}
Combinatorial optimization problems represent a broad class of real-world problems with numerous applications in machine learning, robotics, and bioinformatics. Many such problems are NP-complete or NP-hard, and the computational resources required to solve them on von Neumann machines grow rapidly with problem size. The Ising model represents a promising alternative, where the optimal solution(s) to a combinatorial optimization problem are mapped to the minimum-energy state(s) of a physical system---a network of discrete spins that naturally converges to equilibrium upon perturbation. In this mapping, problem variables correspond to spins, and variable interactions to physical spin-to-spin connections.

Ising machines differ from each other primarily by their inter-spin connectivity and the number of physical spins they support. Densely connected machines can directly express arbitrary variable interactions but support far fewer physical spins under the same hardware budget. Sparsely connected machines accommodate more physical spins \blueHL{but must chain several of them together to represent a single problem variable when a problem needs connections the machine lacks,} yielding a larger and often harder problem to solve. We analyze this tradeoff in detail in Section~\ref{sec:trade-off}.

\begin{figure}[tp]
    \centering
    \subfloat[]{\label{fig:graphs:planarmis}
        \includegraphics[width=.31\linewidth, trim={.8cm .8cm .8cm .8cm}, clip]{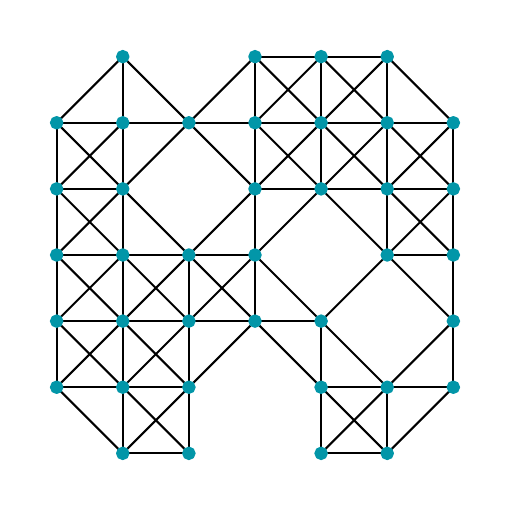}
    }
    \subfloat[]{\label{fig:graphs:sat}
        \includegraphics[width=.31\linewidth, trim={.8cm .8cm .8cm .8cm}, clip]{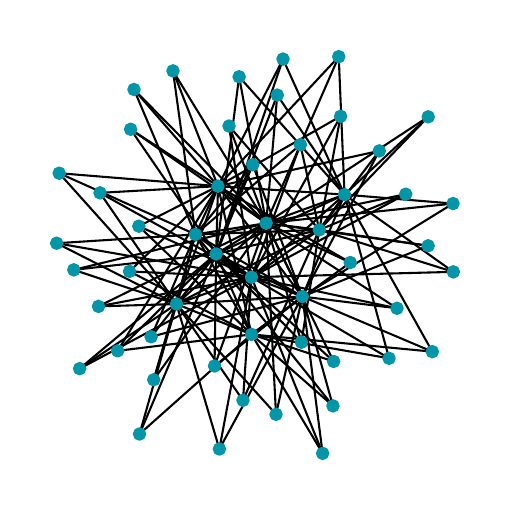}
    }
    \subfloat[]{\label{fig:graphs:nonplanarmis}
        \includegraphics[width=.31\linewidth, trim={.8cm .8cm .8cm .8cm}, clip]{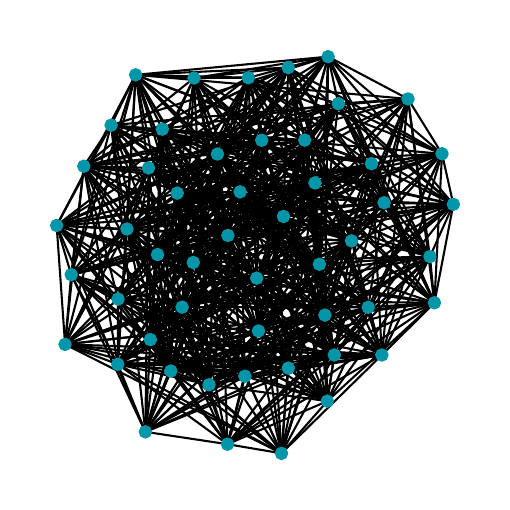}
    }
    \vspace{-0.2cm}
    \caption{Representative problem connectivity graphs: (a) Planar MIS, (b) 3SAT, (c) Nonplanar MIS. Each dot corresponds to a spin, and each edge to a spin-to-spin coupling.}
    \label{fig:pconn}
    \vspace{-0.4cm}
\end{figure}

As the connectivity graphs in Fig.~\ref{fig:pconn} show, combinatorial optimization problems of practical importance exhibit diverse interaction patterns that no single fixed machine connectivity realizes efficiently. Fig.~\ref{fig:graphs:planarmis} shows a planar Maximum Independent Set (MIS) instance, where the goal is to find the largest subset of vertices with no edges between them, with interactions limited to nearest neighbors in a mesh. \blueHL{Fig.~\ref{fig:graphs:sat} shows a 3SAT instance, where the goal is to find Boolean variable assignments that satisfy a given formula. Fig.~\ref{fig:graphs:nonplanarmis} shows a nonplanar MIS instance, where the same problem without the planarity constraint produces far denser interactions.}

At the same time, physical constraints bound both the number of spins and the physical links between them for any given connectivity. \blueHL{These limits vary across technologies, but each admits a maximum physical spin count, $N_{max}$, that is unlikely to keep pace with growing problem sizes.} When the required number of physical spins exceeds machine capacity, the problem must be decomposed into subproblems, which is itself complex because of the data dependencies between them (Section~\ref{sec:background}).

\blueHL{Prior work has scaled Ising machines in three main ways: increasing connectivity toward all-to-all~\cite{Lo2023,cilasun2025cobi}, scaling spin count within a fixed topology on one chip or across many~\cite{moy20221,Kashimata_2024,sharma2022increasing,tatsumura2021scaling,9634769}, or making a single chip reconfigurable~\cite{nikhar2024reconfig}. All of these scale within a single connectivity model. Whether different connectivity classes should instead coexist on one chip is a different question, and it has not been answerable: comparing connectivities at a matched hardware budget would mean fabricating one chip per topology, and physics-accurate simulation cannot reach the scale a workload study requires. This paper introduces \arch: Heterogeneous Ising Multiprocessing, which addresses this mismatch by integrating independent Ising cores of diverse connectivity on one chip and mapping each problem or decomposed subproblem to the core whose connectivity best matches its structure. Under an iso-area budget, this reduces workload completion time by $2.6\times$ on geometric average, and by up to $19\times$, relative to the best homogeneous alternative for each workload.}

\blueHL{In summary, this paper makes the following contributions:}
\begin{list}{\labelitemi}{\leftmargin=1em}
    \item \blueHL{We build the exploration flow that makes the question answerable: a synthesizable Ising emulator that exposes the physical cost of connectivity, and a solver study that establishes what can be run at workload scale.}
    \item \blueHL{We show that no single topology is best across mixed workloads, and that the favorable choice does not follow from a problem's nominal parameters in any simple way.}
    \item \blueHL{We show that a single middle-ground topology does not resolve the mismatch either, which motivates independent cores of fundamentally different connectivity.}
    \item \blueHL{We propose two schedulers that make the problem--core assignment from information a system already has, before or after the Ising mapping, and show that the density-based one recovers essentially the same makespan as an oracle with per-instance knowledge.}
\end{list}

Section~\ref{sec:background} covers background on the Ising model and problem decomposition. Section~\ref{sec:arch} introduces the \arch\ architecture, core independence, and scheduling policies. Section~\ref{sec:eval_setup} describes the emulation framework and evaluation methodology. Section~\ref{sec:eval} presents the quantitative analysis. Section~\ref{sec:related} discusses related work, and Section~\ref{sec:conclusion} concludes.

\section{Background} \label{sec:background}
\noindent{\bf Ising Model}~\cite{v,vi,isingNPHard,viii} models a system of pairwise-interacting spins. A Hamiltonian function captures the energy of an $n$-spin system with $s=\left[ s_1, s_2, ..., s_n \right]$ as:
$$H(s) = -\sum_{<i\neq j>} J_{ij} s_i s_j -\sum_i h_i s_i ~~~ \text{where}~~~i,j \in [1,2, ..., n]$$

\noindent $s_i$ can be either $-1$ or $+1$. $h_i$ denotes the local field at spin $i$, and $J_{ij}$, the interaction strength between spins $i$ and $j$. $h_i$ and $J_{ij}$ are real-valued constants. The system evolves toward low-energy states that minimize $H(s)$ at equilibrium. Positive $J_{ij}$ favors aligned spins, while negative $J_{ij}$ favors anti-aligned spins. The Ising model is isomorphic to Quadratic Unconstrained Binary Optimization (QUBO), given by
$$H(\mathbf{x})=\mathbf{x}^T Q \mathbf{x}~~\text{where}~~ x_i = (s_i + 1)/2$$
for $\mathbf{x} = [x_1, \cdots , x_n]^T \in \{ 0, 1 \}^n$ and $Q \in \mathbb{R}^{n \times n}$.

We distinguish between {logical spins}, which correspond to variables in the Ising/QUBO formulation of a given problem, and {physical spins}, which correspond to the actual hardware implementation. Ideally, each logical spin maps to exactly one physical spin, but limited-connectivity machines may require multiple physical spins to represent a single logical spin (Section~\ref{sec:trade-off}).

\noindent{\bf Problem formulation} maps problem variables to binary-valued spins; problem constraints, to pairwise interactions ($J_{ij}$) and local fields ($h_i$). Many important NP-complete/hard problems, including all of Karp's 21 NP-complete problems, admit Ising formulations~\cite{lucas14}. The goal is for the minimum-energy state(s) to encode the optimal solution(s), which an Ising-compliant physical system can reach through its convergence dynamics~\cite{isingBasicsDWave}.

For a given optimization problem, multiple Ising or QUBO formulations may exist. Because these models primarily capture pairwise interactions, higher-order interactions must be reduced to pairwise ones, often by introducing ancillary variables. \blueHL{This, for instance, is the case for 3SAT, where a Boolean formula of $m$ clauses over $n$ variables must be satisfied. Because each clause is a three-way constraint, the QUBO formulation of~\cite{chancellor2016direct} introduces one ancillary variable per clause, producing an $(m+n)$-variable problem. The mapped problem is therefore larger than the original, and its structure depends on the ratio of clauses to variables.}

\begin{figure}[t]
    \centering
    \includegraphics[width=.25\linewidth]{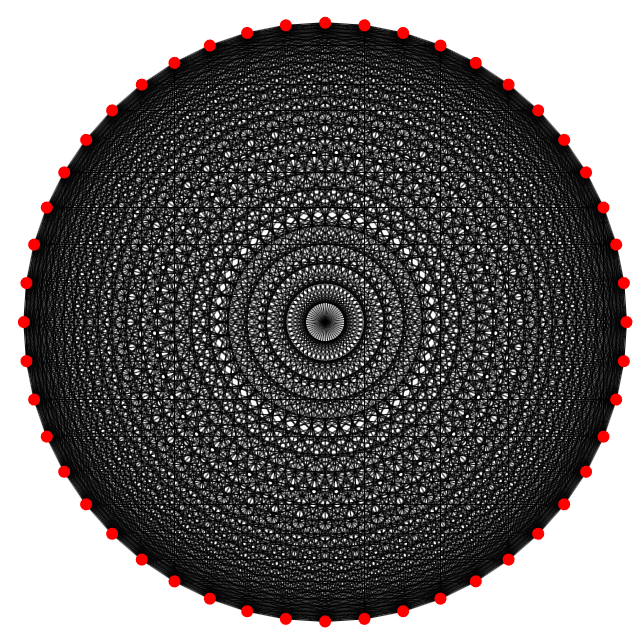}\hspace{0.05\linewidth}%
    \includegraphics[width=.25\linewidth]{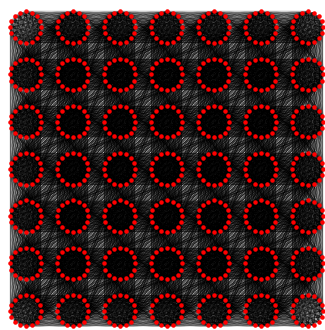}\hspace{0.05\linewidth}%
    \includegraphics[width=.25\linewidth]{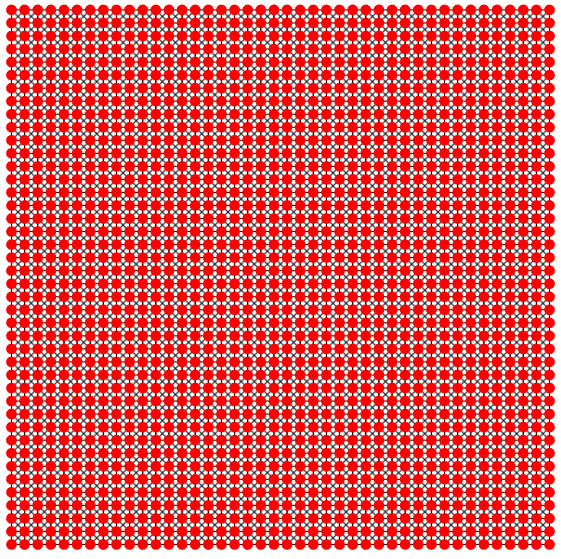}

    \hspace{.19\linewidth}(a)\hfill(b)\hfill(c)\hspace{.2\linewidth}
    \vspace{-0.1cm}
    \caption{Representative Ising machine connectivity:\\
    (a)~All-to-all (A2A), (b)~Hybrid, (c)~King's Graph.}
    \label{fig:conn}
    \vspace{-0.2cm}
\end{figure}

\begin{table}[t]
\small
    \centering
    \begin{tabular}{lccc}
    \toprule
    & A2A & Hybrid & King's \\
    \midrule
    Physical spins ($N_{max}$) & \textbf{48} & \textbf{600} & \textbf{1968} \\
    Area (mm$^2$) & 1.8 & 4.0 & 2.1 \\
    Links per spin & 47 & 111 & 8 \\
    Technology & \multicolumn{3}{c}{65nm CMOS} \\
    Reference & \cite{Lo2023} & \cite{mallick2021overcoming} & \cite{moy20221} \\
    \bottomrule
    \end{tabular}
    \caption{Machine connectivity vs.\ maximum spin count $N_{max}$
    of three representative Ising machines in the same technology.}
    \vspace{-0.4cm}
    \label{tab:conn}
\end{table}

\noindent{\bf Ising Model based solvers} in software and hardware follow different approaches in implementing convergence dynamics. Exhaustive search over all spin assignments scales as $2^n$ and is infeasible for problems of practical size\blueHL{; even at an optimistic 10\,GHz with one configuration evaluated per cycle, a 50-variable problem would require roughly a day of enumeration and a 60-variable problem several years}. Solvers in software therefore rely on probabilistic pruning of the search space, typically using variants of simulated annealing~\cite{sa}, which probabilistically flips selected spin states at each step following predefined convergence criteria. Tabu search~\cite{palubeckis2004multistart} is another general-purpose metaheuristic for QUBO/Ising problems, which iteratively flips spins while maintaining a list of recently visited states to avoid cycling. Solvers in hardware form two broad classes. One class~\cite{takemoto20192,yamamoto2020statica,aramon2019physics,nakayama2021description,goto2021high,tatsumura2021scaling} implements these search procedures directly in hardware. The other directly implements Ising-model-compliant physical systems, spanning quantum~\cite{dwave1,ebadi2022quantum} and quantum-inspired designs~\cite{dutta2021ising,moy20221,Lo2023,cilasun2025cobi,inagaki2016coherent,clements2017gaussian,pierangeli2019large}. In the second class the underlying physical system determines the time evolution of spin states, and the time allocated for convergence is the annealing time. We focus on this class, which we refer to as {\em Ising machines}. \blueHL{A prominent family builds each spin as an electronic oscillator, encoding the spin state in the oscillator's phase and realizing $J_{ij}$ as a coupling between oscillator pairs; we refer to these as oscillator-based Ising machines (OIMs)~\cite{Lo2023,moy20221,cilasun2025cobi}.} Ising machines primarily differ from each other by their physical connectivity\blueHL{, which ranges from sparse nearest-neighbor graphs to all-to-all (Fig.~\ref{fig:conn}). Table~\ref{tab:conn} lists three fabricated examples spanning that range in the same 65nm process}.

\noindent{\bf Problem decomposition} partitions a large problem that exceeds the Ising machine capacity into smaller subproblems that fit within the machine. It becomes necessary as problem sizes grow beyond the spin count supported by a given topology. In the best case, the problem decomposes into independent subproblems that can be solved independently on separate Ising cores. This usually is not possible: subproblems typically depend on each other through shared variables, so the solution to one affects the other, prohibiting subproblem-level parallelism.

For each subproblem, the variables not included must still be accounted for. A common practical approach is {\em clamping}, which sets the corresponding spins in the Ising formulation to known values, typically taken from previously solved subproblems. If, for example, a subproblem covers spin $s_j$, but not $s_i$, by setting $s_i$ to (-)1, the $J_{ij}s_is_j$ term would reduce to (-)$J_{ij}s_j$, which we can account for, without breaking the Ising formulation, by updating the local field coefficient $h_j$ to (-)$J_{ij}$ + $h_j$.

Ising solvers are probabilistic and typically iterate over multiple, possibly overlapping subproblems to cover all problem variables, solving each subproblem a preset number of times. The decomposition algorithm dictates how problem variables get picked in forming subproblems, which has a big impact on solver performance. Divide-and-conquer decomposition is common, with variants that either ignore~\cite{sharma2022increasing} or consider problem connectivity~\cite{cilasun20243sat}.

\section{\arch: Heterogeneous Ising Multiprocessing} \label{sec:arch}
The efficiency of an Ising machine depends on how well its connectivity matches the target problem. We first analyze the tradeoff between connectivity and capacity (Section~\ref{sec:trade-off}), then present the \arch\ architecture (Section~\ref{sec:hetri-arch}) and its workload scheduling policy (Section~\ref{sec:scheduling}).

\subsection{\textbf{The Connectivity-Capacity Tradeoff}}
\label{sec:trade-off}

Implementing the interaction strength $J_{ij}$ requires a physical connection -- a coupler -- between spins $i$ and $j$, and the number of couplers scales fundamentally differently across topologies. An all-to-all (A2A) machine requires $N(N-1)/2$ couplers for $N$ spins, scaling as $O(N^2)$. A King's graph, where each spin connects to at most eight nearest neighbors, requires at most $4N$ couplers, scaling as $O(N)$. Every coupler consumes chip area, wiring tracks, and power. In A2A each additional spin must connect to every existing spin, so the marginal cost of adding a spin grows with $N$; in a King's graph each additional spin connects to a fixed number of neighbors, so the marginal cost stays constant. Under a fixed area budget, $N_{max}$ therefore scales as $\sqrt{\text{budget}}$ for A2A but linearly for King's graph. Table~\ref{tab:conn} shows the consequence in silicon: \blueHL{1968 spins for King's graph versus 48 for A2A at comparable area in the same 65nm process. These are not a controlled comparison -- the designs differ in spin and coupler implementation and in coefficient precision -- but the direction is unambiguous: sparse connectivity buys far more spins under a comparable budget.}

\begin{figure}[tp]
  \centering
  \includegraphics[width=.85\linewidth]{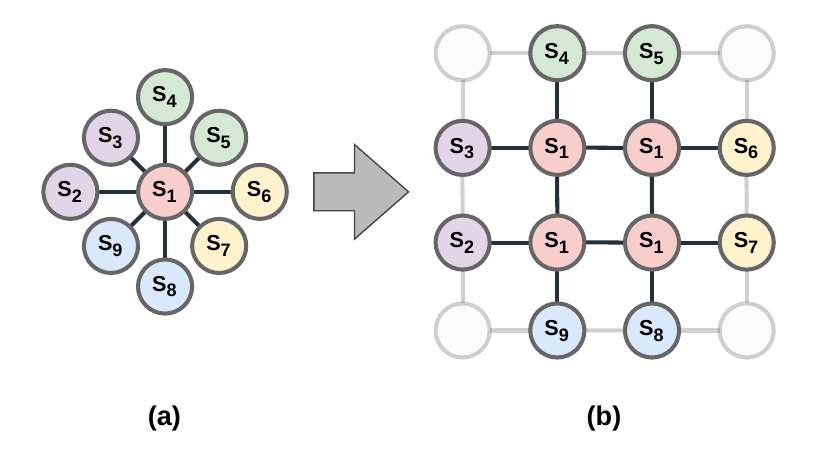}
  \vspace{-0.2cm}
  \caption{Problem connectivity graph capturing problem variable interactions (a), and its embedding onto an Ising machine with limited connectivity (b). Each spin corresponds to a node; each edge in (b), to a physical link. In (b), logical spin $s_1$ requires more connections than a single physical spin can support, so it is replicated across four physical spins, forming a strongly coupled chain.}
  \vspace{-0.4cm}
  \label{fig:embed}
\end{figure}

Sparse topologies avoid the quadratic scaling of all-to-all connectivity, but introduce a different cost: \textbf{embedding}. A2A machines map arbitrary $J_{ij}$ values directly, since a physical link exists between every pair of spins, so embedding is an identity mapping with zero overhead. For any connectivity sparser than A2A, some logical spins require more connections than a single physical spin can support. The solution is to replicate the logical spin across several physical spins forming a strongly coupled chain, as illustrated in Fig.~\ref{fig:embed}. The strong coupling keeps chain members in the same state, while their combined connections serve the actual problem couplings. Finding a valid chain assignment -- which logical spins to replicate, where to place each chain, and how to avoid conflicts between chains competing for the same physical spins -- is a graph minor embedding problem, NP-complete/hard in general~\cite{su2017fast,su2016quantum}.

Beyond the cost of finding an embedding, the duplication carries a \textbf{physical cost}. Each replica consumes a physical spin that could otherwise represent a problem variable, the strong intra-chain coupling slows convergence~\cite{cai2014practical}, and chain coherence becomes harder to maintain as chains grow longer~\cite{hamerly2019experimental,konz2021embedding}. As a representative example, a sparse machine with degree-6 connectivity and $\sim$2000 physical spins can embed a fully connected problem of only $\sim$65 logical variables~\cite{konz2021embedding}, a physical-to-logical ratio exceeding 30:1.

Increasing spin count by reducing connectivity therefore does not translate unconditionally into higher problem capacity. For dense problems, a King's graph may need chains of tens of physical spins per logical spin, spending most of its budget on replication rather than on problem variables. For sparse problems, A2A spends its expensive coupler budget on links the problem never uses, while its limited spin capacity cannot hold even moderately sized instances. \blueHL{The effective capacity of an Ising machine is thus not a fixed property of the hardware, but a function of the problem it is given.}

A middle-ground topology is the natural response. The \textbf{Hybrid graph} of~\cite{mallick2021overcoming}, where spins form all-to-all connected tiles with sparse inter-tile links, scales linearly in couplers like a King's graph but with a larger constant, and retains dense local connectivity. \blueHL{It also inherits limitations from both extremes.} The intra-tile A2A couplers still limit spin count relative to a purely sparse machine, while the sparse inter-tile links still require embedding for problems whose connectivity spans multiple tiles. The tiled structure also introduces a convergence pathology: spins within an A2A tile synchronize rapidly and lock into a collective state that behaves as a single rigid spin, and the sparse inter-tile links lack the coupling strength to perturb these locked clusters, making it hard to escape local minima. As observed experimentally in~\cite{mallick2021overcoming}, this yields fast convergence to a near-optimal solution but requires repeated perturbation to reach the true optimum. \blueHL{The remedy is therefore not a compromise inside one core, but independent cores of fundamentally different connectivity, each operating where it is most effective. This is the insight behind \arch.}

\subsection{\textbf{\arch\ Architecture}}
\label{sec:hetri-arch}

\begin{figure}[t]
    \centering
    \includegraphics[width=\linewidth,trim={0cm 0cm .75cm 0cm},clip]{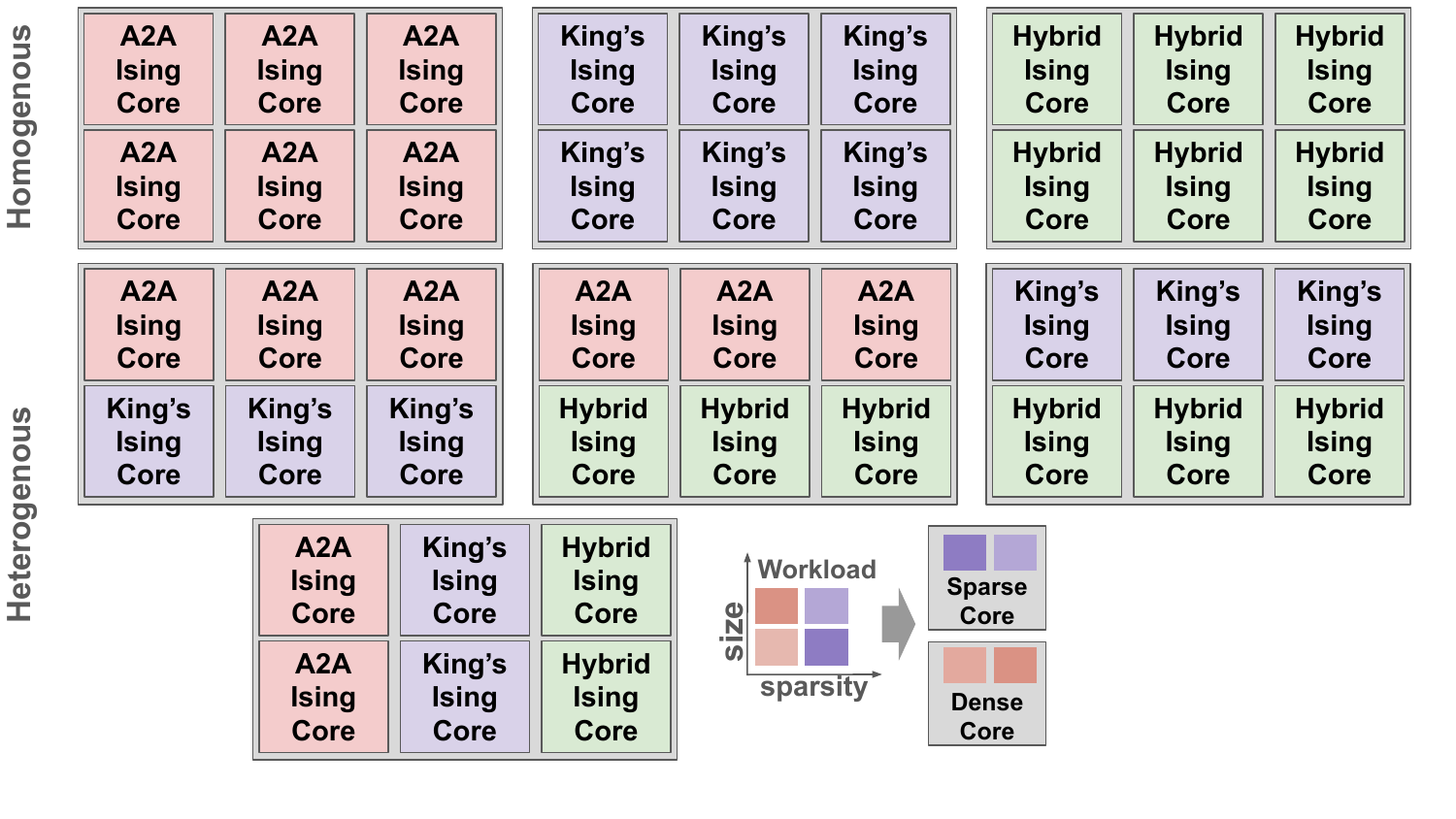}
    \caption{6-core Ising multiprocessor design space with homogeneous baselines and heterogeneous \arch{} core mixes\blueHL{. Fig.~\ref{fig:scheduler_results} evaluates these same design points under different dispatch policies.}}
    \label{fig:het}
    \vspace{-0.6cm}
\end{figure}

\blueHL{\textbf{\arch\ is a heterogeneous multi-core Ising architecture:} a single chip that places several independent Ising cores of different connectivity side by side, with a scheduler that routes each problem to the core whose connectivity best fits it.} We refer to each network of $N_{max}$ spins as an Ising core, where $N_{max}$ is a function of the network topology (Section~\ref{sec:trade-off}). A \arch\ chip features $N_{core}$ such cores, differing in connectivity. Fig.~\ref{fig:het} illustrates example 6-core designs built from A2A, Hybrid, and King's-graph cores. \blueHL{Even with three topology types and six cores, the design space spans three homogeneous and twenty-five heterogeneous configurations, each trading aggregate spin capacity against embedding flexibility and workload coverage; we evaluate representative mixes from this space.} As the number of topology types or cores grows, this space expands combinatorially.

\textbf{\arch\ cores are independent by definition:} there is no physical connection between them. Connecting cores would recreate a Hybrid-like topology at larger scale, reintroducing the problems of Section~\ref{sec:trade-off}. \blueHL{The coupling strength between cores cannot escape this. If it is weak, shared spins drift and the solution breaks; if it is strong enough to hold them together, the coupled cores lock into the same rigid clusters that slow a Hybrid core. Either way, coupling cores rebuilds the middle ground rather than avoiding it.} Instead, \arch\ embraces decomposition. A problem too large for a single core is decomposed into subproblems, each sized to fit one core. Each subproblem is then self-contained, carrying its own spin variables, interaction strengths, and local fields, with the variables it omits accounted for through clamping (Section~\ref{sec:background}). This solve-and-clamp workflow is already standard practice for Ising machines that exceed single-core capacity~\cite{cilasun20243sat,booth2017partitioning}, and \arch's independent cores align with it naturally: each core solves its assigned subproblem without knowledge of the others. \blueHL{There is no shared memory, no inter-core synchronization, and no communication bus; the only coordination happens at the scheduler, which assigns subproblems to cores before computation begins.}

\textbf{The central design question is the composition of the core mix:} for a given $N_{core}$, how many cores of each connectivity type should the chip carry? Sparse problems tend to favor sparse cores that exploit their higher spin capacity, while denser problems favor more directly connected cores. If a workload consisted of a single problem type with known structure, the choice would be clear. In practice workloads are diverse, and different instances within the same family can exhibit substantially different interaction patterns. \blueHL{That diversity persists after decomposition, since neighborhood-based decomposers such as BFS traversal~\cite{cilasun20243sat} preserve much of the original connectivity structure.} Allocating too many cores of one topology wastes silicon when the workload shifts, while allocating too few forces problems onto mismatched cores. \blueHL{No single homogeneous design covers this spectrum, and among heterogeneous mixes the strongest span the connectivity extremes.} We explore this design space quantitatively in Section~\ref{sec:eval}.

The core mix is further constrained by physical realities. Different topologies yield very different spin counts under the same budget, and the differences extend beyond coupler counts to area, power, routing congestion, and programming overhead. These costs are nonlinear, and a topology competitive at small spin counts may be impractical at larger ones. They are not captured by the Ising model and cannot be derived from the connectivity graph alone, which is why we introduce a synthesizable emulator in Section~\ref{sec:emulator}.

\subsection{\textbf{Workload Scheduling}}
\label{sec:scheduling}

\blueHL{A heterogeneous core mix is only useful if each problem can be routed to a core whose connectivity suits it. The cost of a mismatch is not a constant factor: a densely connected instance placed on a sparse core spends its spin budget replicating logical spins into chains rather than representing problem variables (Section~\ref{sec:trade-off}), so the penalty grows with the mismatch. A scheduler makes that routing decision, and the less it needs to know to make it, the more practical the architecture becomes.} \blueHL{We evaluate three dispatch policies over a shared problem pool, each with access to less information than the one before it.}

\blueHL{\textbf{The oracle scheduler} assigns every instance to its individually fastest core, using the true topology-specific TTS of each. It cannot be built, and it establishes whether routing problems to suitable cores is worth doing at all.}

\blueHL{\textbf{The history-based scheduler} tests whether coarse information about the problem is enough. It operates before the Ising mapping, using only the problem as posed: its family and its generating parameter. For each such combination it records the median TTS measured on each topology in past runs and ranks the topologies accordingly. It fits no model; the table stores measured medians and nothing else, and is built once offline.}

\blueHL{\textbf{The density-based scheduler} tests whether the problem needs to be identified at all. It operates after the mapping, computing the coupling density of the delivered QUBO, the fraction of nonzero pairwise couplings, and mapping the instance to a low-, medium-, or high-density regime, each carrying its own topology ranking. No model is fitted: the regime boundaries are read off the measured sweeps of Section~\ref{sec:eval_density}.}

\blueHL{Both proposed policies produce a per-job topology ranking, and both feed the same dispatch loop. Jobs sit in a shared pool; whenever a core becomes free it takes the shortest available job for which its own topology ranks highest, where job length is the median already tabulated for that job's class rather than a per-instance measurement, falling back to lower-ranked jobs if none remain. Homogeneous systems reduce to conventional load balancing, since every core has the same topology and all rankings are equal. Both tables are built from the same instance pool the workloads draw from, but hold only per-class medians, so no job's own solve time affects its dispatch; the reported makespan uses true per-instance times the scheduler never sees.}

\section{Methodology} \label{sec:eval_setup}
Evaluating \arch\ requires \blueHL{knowing what Ising core topologies cost in hardware, and how well they solve problems of diverse structure.} We address the \blueHL{first with} a fully digital, synthesizable Ising-machine emulator (Section~\ref{sec:emulator})\blueHL{, and the second with} an architectural simulation framework (Section~\ref{sec:simulation}), benchmark problems spanning diverse connectivity patterns (Section~\ref{sec:benchmarks}), \arch\ configurations and fairness criteria (Section~\ref{sec:configurations}), and evaluation metrics (Section~\ref{sec:metrics}).

\subsection{\textbf{Digital Ising Emulator}}
\label{sec:emulator}

\begin{figure}[t]
    \centering
    \includegraphics[width=\linewidth,trim={0cm 0cm 0cm 0cm},clip]{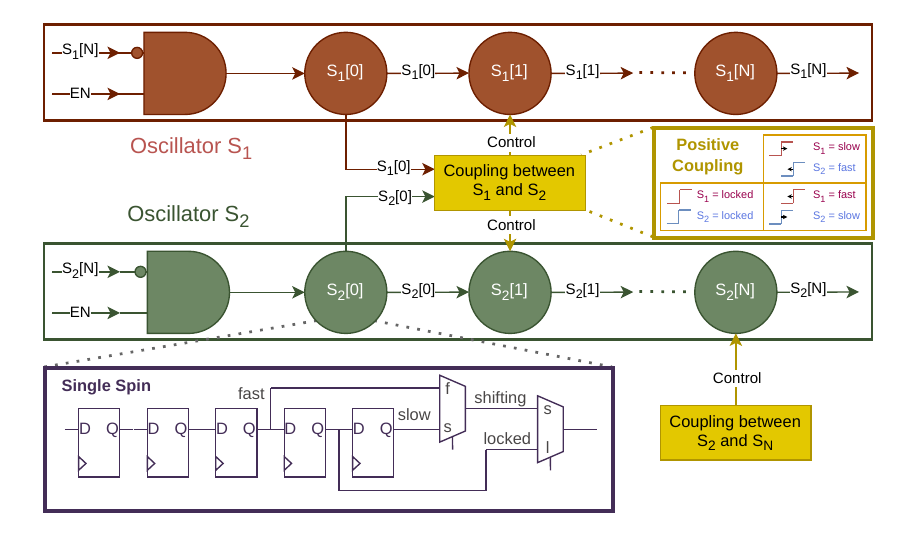}
    \vspace{-.6cm}
    \caption{Digital Ising-machine emulator architecture. Each spin is implemented as a variable-delay element, while peer couplers monitor relative phases and adjust delays to drive spins toward the target coupling.}
    \label{fig:emulation_structure}
    \vspace{-.4cm}
\end{figure}

\blueHL{Isolating the connectivity-capacity tradeoff requires many machines that differ only in connectivity, with spin and coupler resources held equal.} Capturing these effects in analog Ising machines would require full-custom chip design for each topology, which does not lend itself to rapid architectural exploration~\cite{Lo2023,moy20221,cilasun2025cobi}. \blueHL{Existing silicon does not close the gap either. Published machines in the same 65nm CMOS process}~\cite{Lo2023, moy20221, mallick2021overcoming}\blueHL{ confirm that topology governs spin capacity under comparable area (Table~\ref{tab:conn}), but they differ in spin and coupler implementation and in coefficient precision, so their spin counts cannot be compared directly.}

We therefore develop a fully digital, synthesizable \textbf{Ising-machine emulator}. \blueHL{It implements the dynamics of oscillator-based Ising machines (OIMs) and gives every spin and every coupling its own circuit. Unlike digital annealers, which hold coupling coefficients in memory and apply them with shared update logic~\cite{raman2024sachi,yamamoto2020statica}, no coupling here is stored or shared during annealing, so area and routing scale with the problem graph, while the fully digital design can be processed by an automated RTL-to-GDS flow.} \blueHL{Holding the spin and coupler implementation, coefficient precision, and design flow constant while varying only the connectivity is what makes topologies comparable at a matched budget.} It can be parameterized by connectivity type and spin count. Algorithm~\ref{alg:emulator} and Fig.~\ref{fig:emulation_structure} detail the emulator's building blocks.

\blueHL{The emulator and an analog OIM implement the same model. Both follow the Kuramoto dynamics commonly used to describe OIMs~\cite{wang2017oim}, under which each spin's phase evolves as}
\begin{equation}
\dot{\phi}_i = \underbrace{A_c \textstyle\sum_j J_{ij}\,
\tanh\!\big(k\sin(\phi_i-\phi_j)\big)}_{\text{coupling}}
- \underbrace{A_s\sin(2\phi_i)}_{\text{SHIL locking}},
\label{eq:kuramoto}
\end{equation}
\blueHL{where $\phi_i$ is the phase of spin $i$, $J_{ij}$ is the coupling coefficient of Section~\ref{sec:background}, $A_c$ and $A_s$ set the coupling and sub-harmonic injection locking (SHIL) strengths, and $k$ controls how sharply the coupling switches.} In Eq.~\ref{eq:kuramoto}, the coupling term drives each spin toward or away from its neighbor by the sign of their phase difference, which the emulator implements as its $J\oplus\textit{diff}$ decision. The SHIL locking term forces each phase to one of two stable values ($\{0,\pi\}$), which the emulator enforces through the spin's bistable state. Together these mechanisms drive the coupled system toward minima of the Ising Hamiltonian of Section~\ref{sec:background}. \blueHL{We verify this correspondence by simulating a cycle-accurate model of the RTL on random problem instances: the emulator reaches the same solutions as a continuous Kuramoto solver, Tabu search, and a fabricated Ising chip (Section~\ref{sec:simulation}).}

\begin{algorithm}[t]
\caption{Digital Ising Emulator Building Blocks}
\label{alg:emulator}
\begin{algorithmic}[1]
\STATE \textbf{Spin stage} (variable-delay element):
\STATE $Q[0..4] \leftarrow$ 5-stage DFF shift register driven by $\textit{in}$
\STATE $\textit{safe} \leftarrow (Q[2]{=}Q[3]) \wedge (Q[3]{=}Q[4])$
\STATE If \textit{safe}: $\textit{tap} \leftarrow \textit{delay}[1{:}0]$;~ $\textit{out} \leftarrow Q[2{+}\textit{tap}]$ \COMMENT{0=fast, 1=normal, 2=slow}
\STATE \textbf{Peer coupler} ($\textit{in}_1, \textit{in}_2, J, \textit{en} \rightarrow \textit{delay}_1, \textit{delay}_2$):
\STATE $\textit{chg}_1 \leftarrow \textit{in}_1 \oplus \textit{in}_1^{\textit{prev}}$;~~ $\textit{chg}_2 \leftarrow \textit{in}_2 \oplus \textit{in}_2^{\textit{prev}}$;~~ $\textit{diff} \leftarrow \textit{in}_1 \oplus \textit{in}_2$
\IF{$\textit{chg}_1 \wedge \neg\textit{chg}_2$}
    \STATE $\textit{delay}_1 \leftarrow (J \oplus \textit{diff})~?~ \text{fast} : \text{slow}$
\ELSIF{$\textit{chg}_2 \wedge \neg\textit{chg}_1$}
    \STATE $\textit{delay}_2 \leftarrow (J \oplus \textit{diff})~?~ \text{fast} : \text{slow}$
\ELSIF{$\textit{chg}_1 \wedge \textit{chg}_2$}
    \STATE $\textit{delay}_{1,2} \leftarrow (J \oplus \textit{diff})~?~ [\text{normal},\text{normal}] : [\text{slow},\text{fast}]$
\ENDIF
\STATE \textbf{$N$-oscillator A2A core} (illustrated for $N{=}3$):
\STATE Load $J_{ij}$ and enable bits via scan chain
\FOR{each pair $(i,j) \in \{(0,1),(0,2),(1,2)\}$}
    \STATE Coupler$(i,j)$ observes spin$_i$[0], spin$_j$[0] $\rightarrow d_{ij}, d_{ji}$
\ENDFOR
\STATE Delay bus$_i$ = [nominal, $d_{i,\neg i}$]; Oscillator$_i$ = spin-stage chain on bus$_i$
\STATE $\textit{osc\_out}[i] \leftarrow \overline{\textit{spin}_i[3]} \wedge \textit{en}_i$;~ Readout via scan chain
\end{algorithmic}
\end{algorithm}

We instantiate the emulator for A2A, Hybrid, and King's-graph topologies across a range of spin counts using the \textbf{Sky130} open process. The RTL-to-GDS flow is automated with \textbf{LibreLane}~\cite{shalan2020openlane}, while \textbf{OpenROAD}~\cite{ajayi2019openroad} performs placement, routing, and layout generation. For each topology and spin-count configuration, this flow produces a complete backend implementation from which we extract physical-design metrics including area, power, wire length, and cell count.

\begin{figure}[h]
    \centering
    \includegraphics[width=0.7\linewidth,trim={0cm 0cm 0cm 0cm},clip]{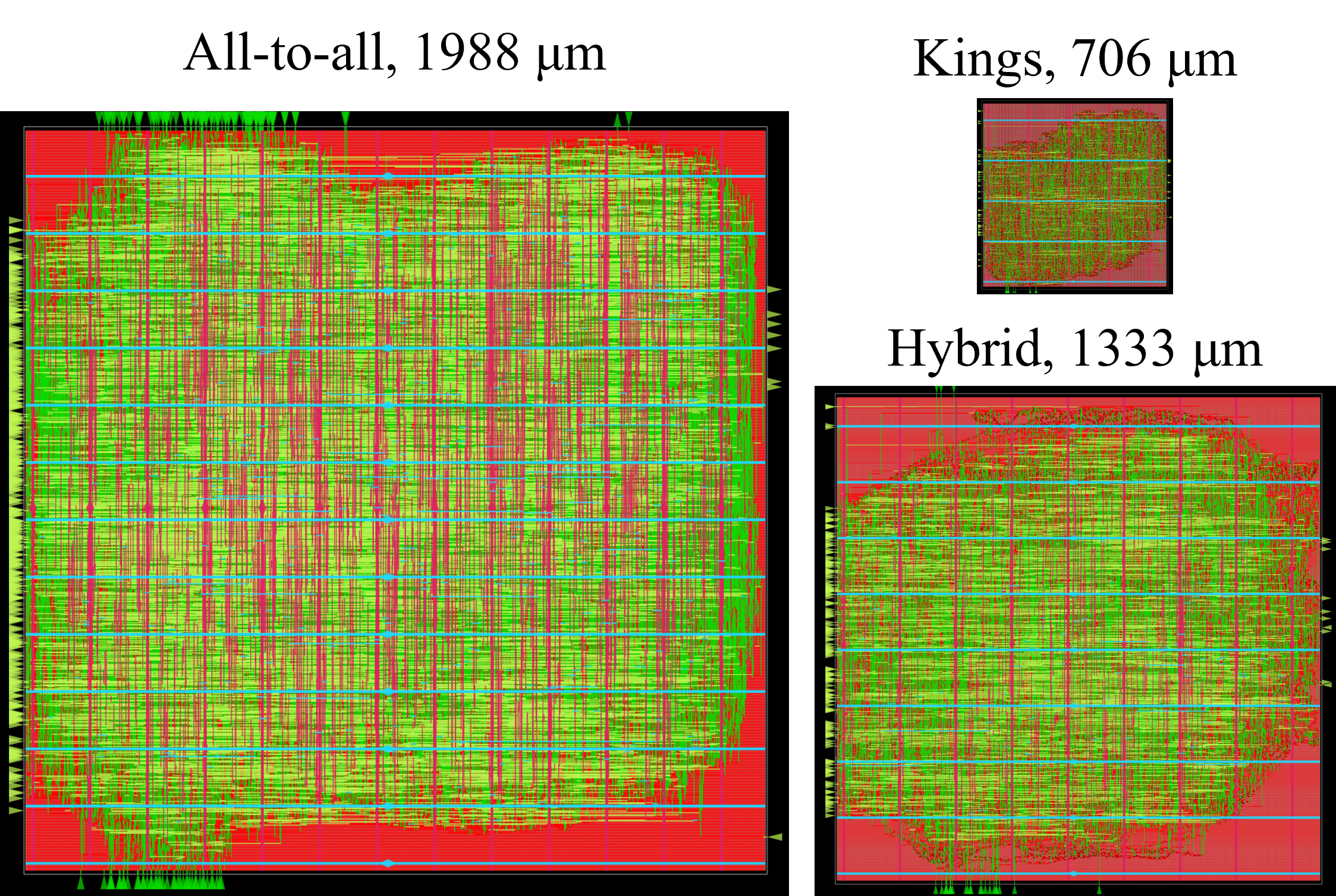}
    \caption{Synthesized Sky130 layouts for emulator-generated 60-spin Ising cores with A2A, Hybrid, and King's-graph topologies.}
    \vspace{-0.4cm}
    \label{fig:chip_layouts_60spin}
\end{figure}

\blueHL{At equal spin count, the topologies occupy visibly different areas (Fig.~\ref{fig:chip_layouts_60spin}).} We use the synthesized backend data to identify iso-area operating points across topologies and to define the fixed hardware configurations used in the evaluation, reported in Section~\ref{sec:eval}.

\subsection{\textbf{Simulation Framework}}
\label{sec:simulation}

While \arch\ applies to any Ising machine technology, we ground our evaluation in oscillator-based Ising machines (OIMs), which represent a leading class of CMOS-compatible Ising hardware with multiple fabricated designs at different connectivity points~\cite{Lo2023,moy20221,mallick2021overcoming}. Evaluating \arch\ requires solving thousands of QUBO instances across different problem sizes, densities, hardware topologies, and decomposition settings.

\begin{figure}[h]
    \centering
    \includegraphics[width=0.7\linewidth]{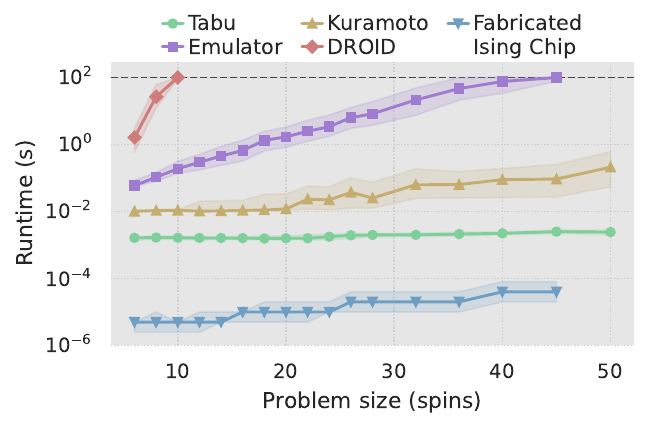}
    \vspace{-.2cm}
    \caption{\blueHL{Time to first reach the reference solution, by problem size. Solver times are CPU wall clock on one machine; the chip figure is measured anneal time and is not comparable to them. Instances are random Ising problems spanning sizes and coupling densities.}}
    \label{fig:solver_validation}
    \vspace{-.2cm}
\end{figure}

\blueHL{\textbf{Two facts shape our choice of solver.} The more of an Ising machine's hardware a tool reproduces, the slower it runs; Fig.~\ref{fig:solver_validation} shows the resulting ordering across a physics-accurate simulator~\cite{kumar2025droid}, our emulator, a continuous Kuramoto integrator, and Tabu search~\cite{palubeckis2004multistart}, all run on the same instances on the same machine. The physics-accurate simulator does not reach the reference solution beyond ten spins within a 100\,s budget, and our emulator saturates at 45 spins. But solving an Ising problem does not require reproducing hardware at all. Tabu search models no Ising machine, reaches the same solutions as the others, and remains fast enough for thousands of instances, so we use it as the solver for the main study.}

\blueHL{The figure also includes a fabricated 28nm CMOS Ising chip, a reference point for what dedicated hardware achieves. It sweeps problem size but not design: one topology, 45 spins. Exploring the space would need a chip per point.}

\blueHL{What the evaluation studies is the consequence of a hardware constraint rather than the behavior of a solver. Each subproblem is restricted to what one core can hold: its physical spin count, and its connectivity after embedding. A problem exceeding that is decomposed into subproblems that fit, and each round issues one such subproblem to one core and receives one solution back. Any engine that solves at core size can serve that call, which is what Fig.~\ref{fig:solver_validation} shows. Prior work has also shown that Tabu-search-based software workflows closely track the iteration-level behavior of fabricated oscillator-based Ising chips on the same problem instances~\cite{cilasun20243sat}. The round count that Section~\ref{sec:metrics} converts into core time therefore reflects the mapping---embedding overhead, effective problem size increase due to spin replication inflation, and connectivity mismatch---rather than the engine that served the call.}

\textbf{For non-A2A cores,} each problem is first embedded onto the target connectivity graph via graph minor embedding, inflating the problem by replicating logical spins across physical chains (Section~\ref{sec:trade-off}). The embedded problem is then solved by Tabu search, and the resulting \blueHL{number of solve-and-clamp rounds} reflects the difficulty of the inflated problem on that topology. \blueHL{Tabu reaches these solutions within one to two milliseconds across the sizes evaluated, so each subproblem solve is allowed three milliseconds. A much longer 5000~ms run provides the reference energy against which all topologies are compared (Section~\ref{sec:metrics}).}

\textbf{For problem decomposition,} we adapt the BFS-based decomposer from~\cite{cilasun20243sat}. Starting from a random root node, the decomposer traverses the problem connectivity graph via BFS to form an ordered local neighborhood of variables. Variables outside the selected neighborhood are clamped to their best-known values, folding their contributions into the local field coefficients of the remaining variables (Section~\ref{sec:background}). \blueHL{The scheduling policy of Section~\ref{sec:scheduling} runs on top of this framework for both homogeneous and heterogeneous \arch\ configurations.}

\subsection{\textbf{Benchmarks}}
\label{sec:benchmarks}

\begin{figure}[t]
    \centering
    \captionsetup[subfloat]{captionskip=1pt,farskip=0pt}
    \subfloat[$p=0.04$]{\label{fig:adj:er0_04}%
    \includegraphics[width=0.31\linewidth,trim={0cm 0cm 0cm 0cm},clip]{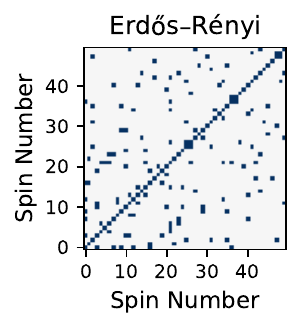}}\hfil
    \subfloat[$p=0.16$]{\label{fig:adj:er0_16}%
    \includegraphics[width=0.31\linewidth,trim={0cm 0cm 0cm 0cm},clip]{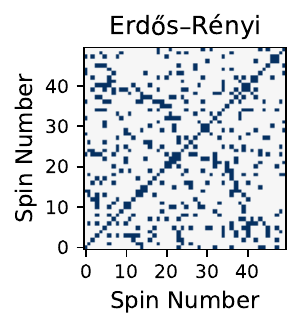}}\hfil
    \subfloat[$p=0.64$]{\label{fig:adj:er0_64}%
    \includegraphics[width=0.31\linewidth,trim={0cm 0cm 0cm 0cm},clip]{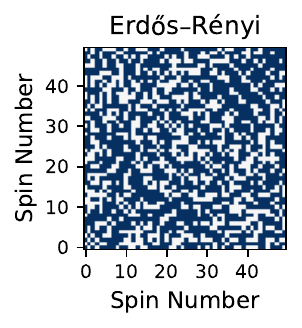}}\\[-2pt]
    \subfloat[$m=1$]{\label{fig:adj:ba1}%
    \includegraphics[width=0.31\linewidth,trim={0cm 0cm 0cm 0cm},clip]{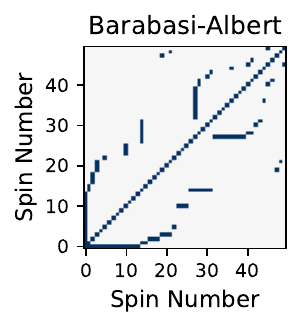}}\hfil
    \subfloat[$m=2$]{\label{fig:adj:ba2}%
    \includegraphics[width=0.31\linewidth,trim={0cm 0cm 0cm 0cm},clip]{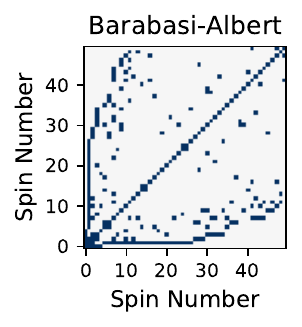}}\hfil
    \subfloat[$m=8$]{\label{fig:adj:ba8}%
    \includegraphics[width=0.31\linewidth,trim={0cm 0cm 0cm 0cm},clip]{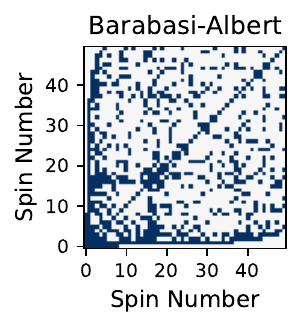}}\\[-2pt]
    \subfloat[$m=1$]{\label{fig:adj:pow1}%
    \includegraphics[width=0.31\linewidth,trim={0cm 0cm 0cm 0cm},clip]{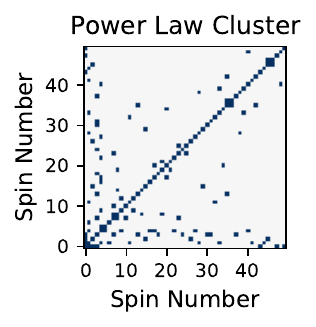}}\hfil
    \subfloat[$m=4$]{\label{fig:adj:pow4}%
    \includegraphics[width=0.31\linewidth,trim={0cm 0cm 0cm 0cm},clip]{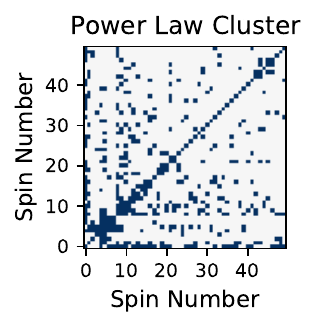}}\hfil
    \subfloat[$m=16$]{\label{fig:adj:pow16}%
    \includegraphics[width=0.31\linewidth,trim={0cm 0cm 0cm 0cm},clip]{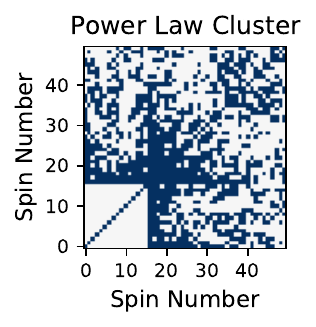}}\\[-2pt]
    \subfloat[$\alpha=1$]{\label{fig:adj:sat1}%
    \includegraphics[width=0.31\linewidth,trim={0cm 0cm 0cm 0cm},clip]{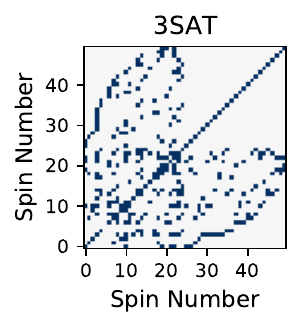}}\hfil
    \subfloat[$\alpha=2$]{\label{fig:adj:sat2}%
    \includegraphics[width=0.31\linewidth,trim={0cm 0cm 0cm 0cm},clip]{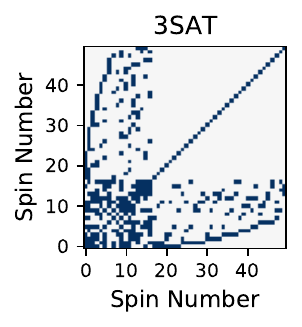}}\hfil
    \subfloat[$\alpha=5$]{\label{fig:adj:sat5}%
    \includegraphics[width=0.31\linewidth,trim={0cm 0cm 0cm 0cm},clip]{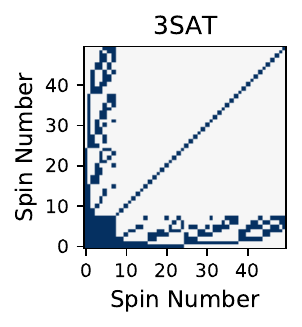}}
    \vspace{-.1cm}
    \caption{Representative benchmark connectivity patterns, shown as adjacency matrices, one family per row: (a)--(c) Erd\H{o}s--R\'enyi, (d)--(f) Barab\'asi--Albert, (g)--(i) Power-Law Cluster, and (j)--(l) SAT-derived QUBO instances.}
    \label{fig:adj}
    \vspace{-.3cm}
\end{figure}

We evaluate \arch\ using benchmark families that span a broad range of connectivity patterns and problem classes, including graph-based QUBO instances and SAT-derived QUBO instances. To enable controlled comparison across families, we normalize all constructed QUBO instances to 50 logical variables. \blueHL{This size balances two constraints. Validating solver behavior against physics-level models becomes impractical beyond a few tens of spins (Fig.~\ref{fig:solver_validation}), while a 50-variable problem already expands to several hundred physical spins once embedded onto a sparse topology.}

We consider three graph-based QUBO instance families. \textbf{Erd\H{o}s--R\'enyi} (\texttt{er}) graphs~\cite{bothers} approximate homogeneous random interaction patterns~\cite{newman2003structure,erdos1959random}, and we sweep graph density from 0.01 to 0.64. \textbf{Barab\'asi--Albert} (\texttt{ba}) graphs~\cite{bothers} model growing networks with preferential attachment~\cite{barabasi1999emergence,newman2003structure}, and we sweep the attachment parameter $m$ from 1 to 32. \textbf{Power-Law Cluster} (\texttt{pow}) graphs~\cite{holme2002growing} capture heavy-tailed connectivity with clustering~\cite{eikmeier2017revisiting,newman2003structure}; we fix the triad-formation probability to 0.5 and sweep $m$ from 1 to 16, with an additional dense point at $m=32$. For \textbf{SAT-derived QUBO} benchmarks (\texttt{sat}), we use the formulation from~\cite{chancellor2016direct} and sweep the clause-to-variable ratio $\alpha$ from 0.5 to 5.0 in increments of 0.25. Across these families, we generate several thousand random problem instances. Fig.~\ref{fig:adj} shows representative adjacency matrices spanning this range.

From these families we construct \blueHL{15 workloads, one for every non-empty combination of the four families: the four single-family workloads, six pairs, four triples, and one containing all four.} This lets us evaluate topology matching both within a single benchmark class and across mixed workloads, and later assess how effectively the scheduler exploits topology diversity.

\subsection{\textbf{Compared Configurations and Fairness Criteria}}
\label{sec:configurations}

This paper is not about the design of a specific Ising core, but about an architectural principle that applies across Ising core implementations. Any design can serve as a component core of \arch, provided it uses the same underlying technology while exposing substantially different connectivity. We base our evaluation on three representative oscillator-based organizations that satisfy these criteria: A2A, Hybrid, and King's graph~\cite{Lo2023,mallick2021overcoming,moy20221}, \blueHL{each demonstrated in silicon (Table~\ref{tab:conn}). These three bound the tradeoff rather than sample it. A2A and King's graph sit at opposite ends of the connectivity--capacity range, the densest practical connectivity and one of the sparsest, so they bracket the regime the design question lives in. Hybrid is included as a control rather than as a third design point: it tests whether a single compromise topology can achieve what \arch\ achieves with independent cores, and Section~\ref{sec:eval_density} shows that it cannot.}

Our primary fairness criterion is \textbf{iso-area} comparison. Rather than forcing all topologies to use the same spin count, we match them by hardware budget using the synthesized backend data from the emulator, so that performance differences reflect topology rather than unequal area allocation. \blueHL{We additionally check total power at the matched points, so that an area match does not conceal a power imbalance.} The specific operating points \blueHL{and the resulting power spread} are reported in Section~\ref{sec:eval}.

At the system level, we compare both homogeneous and heterogeneous \arch\ organizations. This lets us evaluate not only which topology is favorable for a given problem regime, but also whether mixing complementary topologies improves coverage across diverse workloads.

\subsection{\textbf{Metrics}}
\label{sec:metrics}

We evaluate \arch\ using time-to-solution (TTS), which captures how quickly a hardware configuration reaches a target solution. Because the solver is stochastic, we evaluate each problem--hardware pair over 128 repeated trials and aggregate the outcomes.

Physical annealing time cannot be measured directly from the RTL, and how long an Ising machine must anneal to reach the reference solution is not predictable in advance, so we model core time with a topology-dependent proxy: each solve-and-clamp round is charged one traversal of the core's oscillator chain. For each topology,
\begin{equation}
T_{\mathrm{iter}} = N_{\mathrm{stages}} \times T_{\mathrm{clk}},
\label{eq:titer}
\end{equation}
where $N_{\mathrm{stages}}$ is the number of variable-delay elements forming a spin's oscillator in the synthesized core and $T_{\mathrm{clk}}$ is the clock period, \blueHL{since a spin advances one stage per clock cycle. A spin needs one delay element per coupling it participates in, plus a nominal element (Algorithm~\ref{alg:emulator}), so $N_{\mathrm{stages}}$ follows the topology's degree rather than its spin count. All three cores are synthesized at a common period of $T_{\mathrm{clk}} = 50$\,ns, which is the value used throughout. Coefficient programming and spin readout are handled by parallel scan chains of fixed depth, whose area and power cost is already carried by the synthesized designs; their latency is a constant independent of spin count and topology, so it does not enter the per-iteration model.} \blueHL{Rounds} are converted to core time as
\begin{equation}
T_{\mathrm{core}} = N_{\mathrm{rounds}} \times T_{\mathrm{iter}}.
\label{eq:tchip}
\end{equation}
\blueHL{Because $T_{\mathrm{clk}}$ is common to all three topologies, it scales every configuration equally and the relative TTS rankings that drive our conclusions do not depend on its exact value.}

Exhaustive enumeration of a 50-variable instance would take about a day (Section~\ref{sec:background}), and we generate several thousand, so we define a reference target energy for each using the same Tabu engine with a much longer timeout of 5000 milliseconds. Topology-specific runs then use the 3 millisecond timeout of Section~\ref{sec:simulation}, and configurations are compared by how quickly they reach this common reference. All topologies are therefore measured against the same target, and differences reflect speed rather than solution quality.

Given a target success probability $p_{\mathrm{target}}$, we compute TTS in the standard form~\cite{Boixo_2014}
\begin{equation}
\mathrm{TTS}(t) = t \cdot \frac{\log(1-p_{\mathrm{target}})}{\log(1-p(t))},
\label{eq:tts}
\end{equation}
where \blueHL{$t = T_{\mathrm{core}}$} is the core completion time of a trial and $p(t)$ is the empirical probability of reaching the reference energy within time $t$. We set $p_{\mathrm{target}}=0.95$ and report the minimum of Eq.~\ref{eq:tts} over the observed completion times.

\blueHL{A workload is a set of problem instances dispatched across a configuration's cores by the scheduler of Section~\ref{sec:scheduling}. Each instance occupies its assigned core for its TTS, and we report the workload makespan: the time at which the last instance completes. All \arch\ results in Section~\ref{sec:eval_scheduler} are makespans in this sense. Reported times are chip-time projections from Eqs.~\ref{eq:titer}--\ref{eq:tts}: host-side work, including embedding, decomposition, and scheduling, runs outside the cores and is not counted. That omission is not neutral, since only the non-A2A topologies embed at all, and we therefore compare cores rather than complete systems.}

\section{Evaluation} \label{sec:eval}
\begin{figure}[h]
    \centering
    \includegraphics[width=0.8\linewidth,trim={0.7cm 0.3cm 0.7cm 0.5cm},clip]{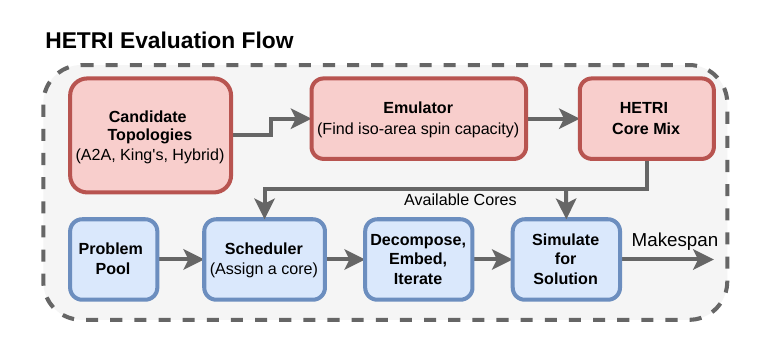}
    \vspace{-0.1cm}
    \caption{\blueHL{Evaluation flow. The emulator fixes each topology's spin count at a matched area budget (Section~\ref{sec:emulator}); the scheduler assigns problems to cores (Section~\ref{sec:scheduling}); each problem is decomposed, embedded, and solved (Section~\ref{sec:simulation}).}}
    \label{fig:overview}
    \vspace{-0.4cm}
\end{figure}

We evaluate \arch\ along three dimensions, following the flow of Fig.~\ref{fig:overview}: hardware cost, topology matching, and scheduler-driven workload performance.

\subsection{\textbf{Emulator-Based Hardware Analysis}}
\label{sec:eval_hardware}

\begin{figure}[h]
    \centering
    \subfloat[Core area vs.\ spin count\label{fig:chip_ppa:area}]{
        \includegraphics[width=0.47\linewidth,trim={0cm 0cm 0cm 0cm},clip]{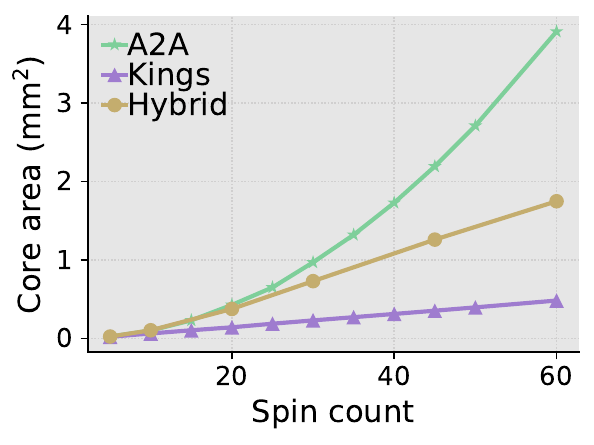}
    }
    \hfill
    \subfloat[Total power vs.\ spin count\label{fig:chip_ppa:power}]{
        \includegraphics[width=0.47\linewidth,trim={0cm 0cm 0cm 0cm},clip]{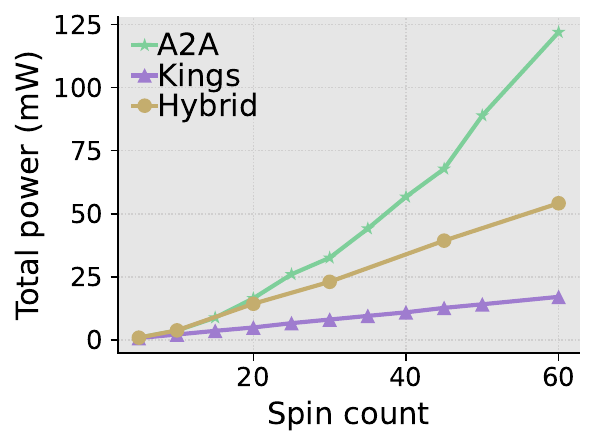}
    }
    \caption{Synthesized backend scaling of emulator-generated Ising cores: (a) core area and (b) total power versus spin count. Each point is a distinct synthesized design produced by the same RTL-to-GDS flow.}
    \label{fig:chip_ppa}
    \vspace{-0.4cm}
\end{figure}

\blueHL{Fig.~\ref{fig:chip_ppa} shows how backend cost scales with spin count for the three topologies synthesized with the flow of Section~\ref{sec:emulator}. Cost tracks each topology's coupler count: A2A area grows quadratically with spin count ($\sim\!N^{2}$), King's graph near-linearly ($\sim\!N$), and Hybrid in between, matching the tradeoff of Section~\ref{sec:trade-off}.}

Table~\ref{tab:hetri_backend_isoarea} lists the iso-area operating points selected from these sweeps. At approximately the same core area, the three topologies expose markedly different spin capacities: \textbf{34} spins for A2A, \textbf{45} for Hybrid, and \textbf{144} for King's graph. \blueHL{The matched points also lie within roughly 17\% of one another in total power (39--46\,mW), so the comparison is close to \textbf{iso-power} as well, and the workload differences that follow reflect topology rather than silicon or power allocation.}

\begin{table}[H]
\centering
\small
\caption{Backend comparison of iso-area operating points across A2A, Hybrid, and King's-graph Ising cores synthesized in Sky130.}
\label{tab:hetri_backend_isoarea}
\begin{tabular}{lccc}
\toprule
\textbf{Metric} & \textbf{A2A} & \textbf{Hybrid} & \textbf{King's} \\
\midrule
Spins                   & \textbf{34}  & \textbf{45}  & \textbf{144} \\
Core area (mm$^2$)      & 1.25 & 1.26 & 1.23 \\
Total power (mW)        & 41.79 & 39.31 & 46.10 \\
Wirelength (mm)         & 1635 & 1477 & 1440 \\
Total cells (k)     & 70.7 & 71.6 & 68.8 \\
\bottomrule
\end{tabular}
\vspace{-.3cm}
\end{table}

These operating points also fix the per-iteration core-time proxy of Eq.~\ref{eq:titer}. \blueHL{Because a spin needs one delay element per coupling, stage count follows degree rather than spin count: the 34-spin A2A design needs 34 stages and the 45-spin Hybrid 35, while the 144-spin King's-graph design needs only 9. At the 50\,ns clock period this yields per-iteration times of 1700\,ns, 1750\,ns, and 450\,ns. The sparse core is therefore both larger in capacity and faster per iteration. At this spin count the Hybrid core's tiles leave it denser relative to its size than a larger fabricated Hybrid would be, while still requiring embedding for couplings that span tiles.} We use these values to convert \blueHL{round counts} into core-level time-to-solution.

\subsection{\textbf{Problem Density and Topology Matching}}
\label{sec:eval_density}

Having fixed the hardware budget through Table~\ref{tab:hetri_backend_isoarea}, we evaluate the three cores across sweeps spanning a broad range of connectivity regimes. The question is \textbf{which topology is favorable for which class of problems} under the same silicon budget, and whether any homogeneous design remains best across all of them.

\begin{figure}[H]
    \centering
    \captionsetup[subfloat]{captionskip=1pt,farskip=0pt}
    \subfloat[Erd\H{o}s--R\'enyi\label{fig:density_match:er}]{%
    \includegraphics[width=0.8\linewidth,trim={0.2cm 0.2cm 0.2cm 0.2cm},clip]{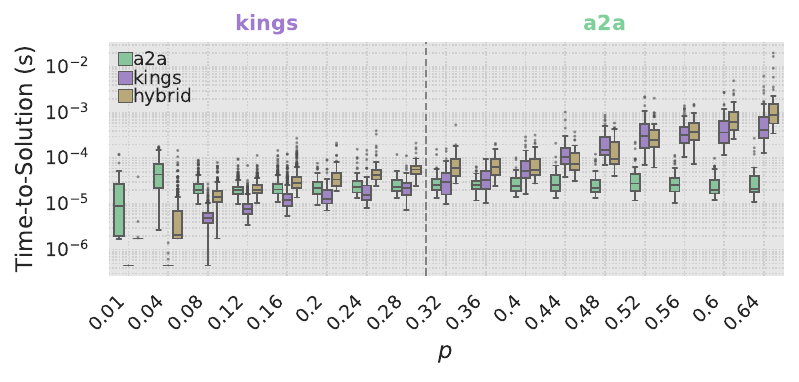}}\\[-1pt]
    \subfloat[Barab\'asi--Albert\label{fig:density_match:ba}]{%
    \includegraphics[width=0.8\linewidth,trim={0.2cm 0.2cm 0.2cm 0.2cm},clip]{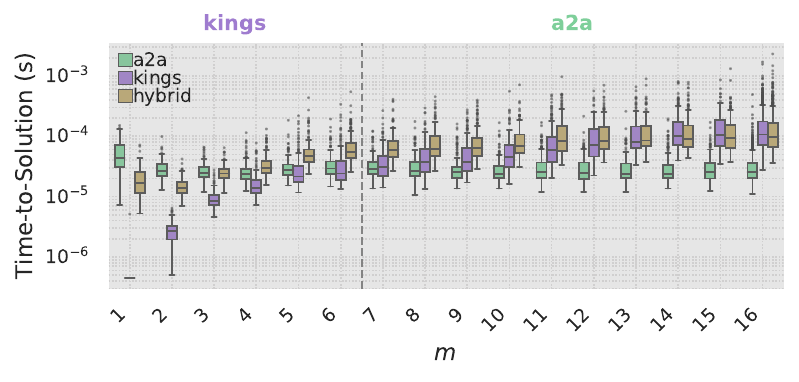}}\\[-1pt]
    \subfloat[Power-Law Cluster\label{fig:density_match:pr}]{%
    \includegraphics[width=0.8\linewidth,trim={0.2cm 0.2cm 0.2cm 0.2cm},clip]{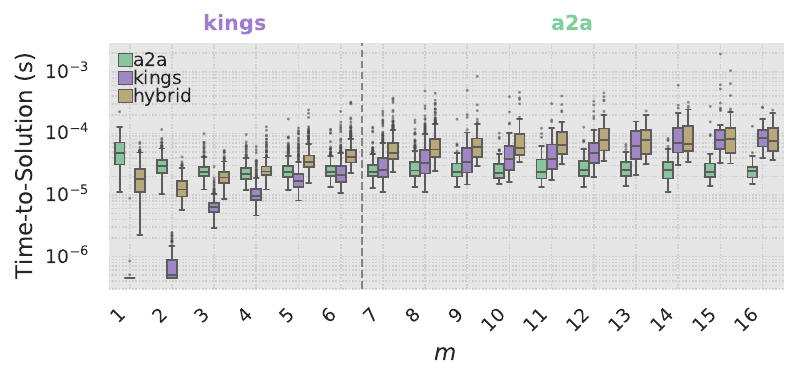}}\\[-1pt]
    \subfloat[SAT-derived QUBO\label{fig:density_match:sat}]{%
    \includegraphics[width=0.8\linewidth,trim={0.2cm 0.2cm 0.2cm 0.2cm},clip]{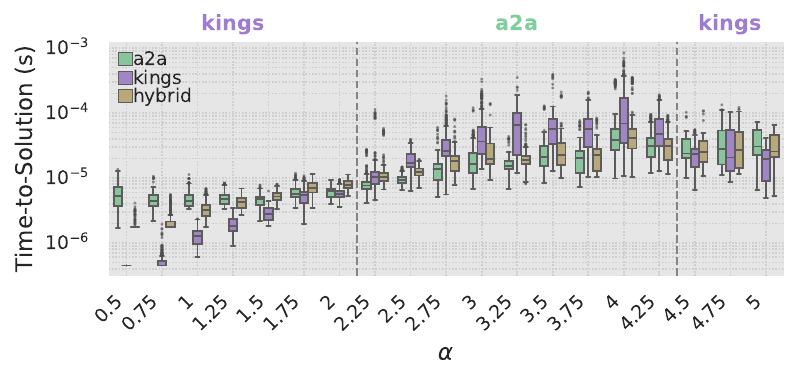}}
    \vspace{-0.1cm}
    \caption{Topology-dependent core time-to-solution across benchmark density sweeps for the iso-area A2A, Hybrid, and King's-graph cores. Each box shows the distribution over repeated trials; annotations mark the topology with the lowest median TTS in each regime.}
    \label{fig:density_match}
    \vspace{-0.45cm}
\end{figure}

Fig.~\ref{fig:density_match} shows that topology preference changes systematically with problem structure. \blueHL{Erd\H{o}s--R\'enyi instances (a)} strongly favor the King's-graph core \blueHL{while the graph is sparse, with the advantage shifting to A2A as density rises; the crossover falls later here than for the other graph families. Barab\'asi--Albert (b) and Power-Law Cluster (c)} show the same one-way pattern, with the shift occurring once connectivity grows beyond the low-$m$ regime.

\begin{figure*}[tp]
\centering
\subfloat[Oracle scheduler\label{fig:sched_ideal}]{
\includegraphics[width=0.80\linewidth,trim={0.5cm 0.5cm 0.5cm 0.5cm},clip]{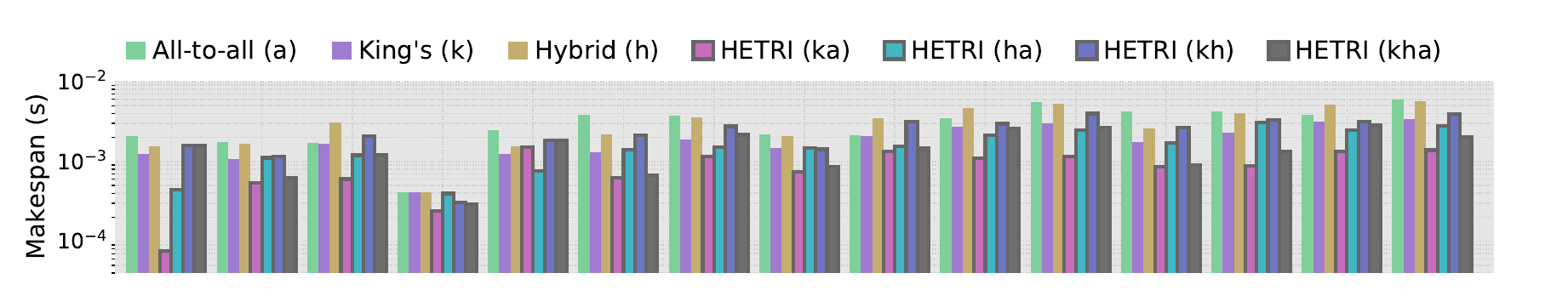}
}\
\subfloat[History-based scheduler\label{fig:sched_history}]
{
\includegraphics[width=0.80\linewidth,trim={0.5cm 0.4cm 0.5cm 0.1cm},clip]{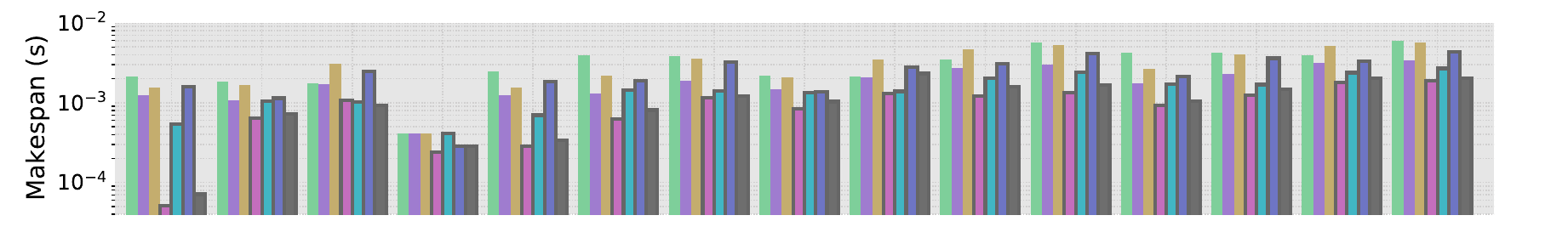}
}\
\subfloat[Density-based scheduler\label{fig:sched_density}]
{
\includegraphics[width=0.80\linewidth,trim={0.5cm 1.2cm 0.5cm 0.2cm},clip]{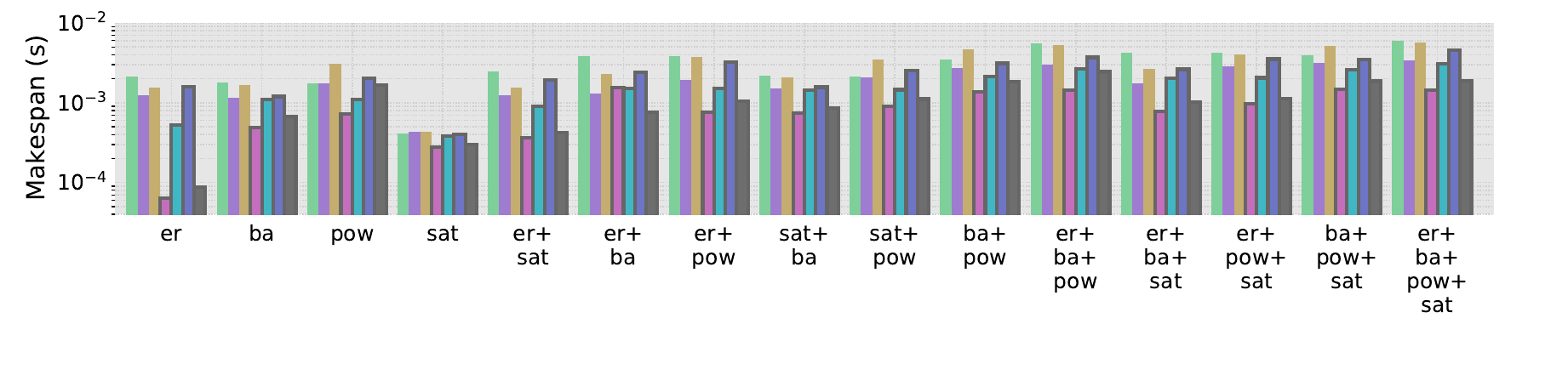}
}
\caption{Average workload makespan across homogeneous baselines and heterogeneous \arch{} configurations drawn from the six-core design space of Fig.~\ref{fig:het}, under three scheduling policies. In the legend, a, h, and k denote All-to-all, Hybrid, and King's-graph cores; two-letter \arch{} configurations use three cores of each listed topology, while \arch\ (kha) uses two of each. Lower makespan is better.}
\label{fig:scheduler_results}
\vspace{-.3cm}
\end{figure*}

\blueHL{In these three families the homogeneous Hybrid core almost never attains the lowest median TTS. A middle-ground topology therefore does not resolve the connectivity tradeoff. Rather than combining the strengths of both extremes, Hybrid inherits A2A's capacity limit and a sparse graph's embedding burden. Capacity and embedding alone account for the result, without invoking the convergence pathology of Section~\ref{sec:trade-off}.}

SAT-derived instances behave differently \blueHL{(Fig.~\ref{fig:density_match}(d))}. King's graph is favorable in the sparse regime, A2A takes over across intermediate clause densities, and King's graph becomes competitive again at high $\alpha$. \blueHL{This reversal shows that topology preference is not set by the nominal sweep parameter, but by how the SAT-to-QUBO transformation shapes the resulting interaction graph.}

\blueHL{\textbf{The mapped QUBOs explain why.} Characterizing the generated instances directly, the three graph families rise monotonically in QUBO interaction density across their range, reaching 0.64 for Erd\H{o}s--R\'enyi and 0.43--0.44 for Barab\'asi--Albert and Power-Law Cluster, matching their one-way crossover. SAT-derived QUBOs instead span a much narrower band and turn over: density rises from 0.08 at $\alpha=0.5$ to a peak of 0.14 near $\alpha=2.5$, then falls back to 0.13 by $\alpha=5$, while the maximum variable degree keeps climbing from 13 to 27. High-$\alpha$ instances therefore become degree-skewed rather than uniformly denser, and the point where King's graph regains the lead coincides with that downturn.}

\blueHL{No single homogeneous topology therefore covers all families and density regimes. The favorable choice tracks the structure of the mapped interaction graph rather than the nominal sweep parameter, and that structure is visible in the delivered QUBO itself. Section~\ref{sec:eval_scheduler} shows it can be turned into a dispatch rule.}

\subsection{\textbf{Scheduler-Aware System Evaluation}}
\label{sec:eval_scheduler}

We compare homogeneous and heterogeneous 6-core \arch\ organizations under the three dispatch policies of Section~\ref{sec:scheduling}, shown in Fig.~\ref{fig:scheduler_results}. \blueHL{The question here is not how much benefit a scheduler can extract, but whether the assignment \arch\ depends on can be made at all: the oracle establishes that core choice is worth making, and the two proposed schedulers establish that it can be made from information a system actually has.}

\blueHL{Under all three policies and on every workload, the lowest makespan is achieved by a heterogeneous organization; no homogeneous A2A-, Hybrid-, or King's-only system is ever the best of the evaluated configurations. Heterogeneity alone is not sufficient, however. The King's+Hybrid mix (\arch-kh) often trails the strongest configurations and at times only matches the homogeneous cores. The strongest configurations pair the connectivity extremes: \arch-ka, spanning the sparse and dense ends of the range, is the best of the evaluated configurations on nearly all workloads.}

\blueHL{Under the oracle (Fig.~\ref{fig:scheduler_results}(a)),} \arch-ka delivers the lowest makespan by a wide margin on \texttt{er}, \texttt{ba}, and \texttt{pow}; on \texttt{er} it reaches $7.60\times10^{-5}$\,s against $2.07\times10^{-3}$\,s for A2A-only and $1.24\times10^{-3}$\,s for King's-only. \blueHL{The preferred mix can shift with workload composition: on the mixed \texttt{er+sat} workload the Hybrid+A2A configuration (\arch-ha) is fastest at $7.64\times10^{-4}$\,s, ahead of \arch-ka at $1.51\times10^{-3}$\,s. A Hybrid core is rarely the best topology alone, yet it can contribute to the best mix when paired with A2A on a workload containing both sparse and dense structure.}

\blueHL{The history-based scheduler in Fig.~\ref{fig:scheduler_results}(b) achieves the same benefit using only the problem's family and parameter, known before any Ising mapping takes place. Heterogeneous organizations again attain the lowest makespan on every workload, and \arch-ka is best on all but \texttt{pow}, where \arch-kha edges ahead with \arch-ka close behind. The preference for one core over another is therefore already present at the problem level, and a lookup table built once from past runs is enough to exploit it.}

\blueHL{The density-based scheduler in Fig.~\ref{fig:scheduler_results}(c) achieves it again with no family label and no history at all, partitioning the density range into preference zones that order the cores, with a narrow band between zones where the top two are nearly tied. Hybrid is the hardest core for a density rule to place and usually falls in that band, but this costs little, since Hybrid is rarely the optimal target and the zone boundaries sit around the King's--A2A crossover where makespan is actually determined. Across all 15 workloads it beats every homogeneous baseline, by $2.6\times$ on geometric average and up to $19\times$. Averaged over all workloads the oracle, history-based, and density-based policies reduce makespan relative to the best homogeneous baseline by $2.4\times$, $2.3\times$, and $2.6\times$ respectively, all within roughly 10\% of one another. The oracle is not makespan-optimal: it assigns each instance to its individually fastest core, which can leave that core's queue imbalanced, so a load-balancing policy can occasionally finish sooner while knowing less. The signal that makes the assignment possible therefore survives the mapping to QUBO, and can be read from the coupling matrix of a problem whose origin is unknown.}

\blueHL{\textbf{The precondition \arch\ depends on is therefore met at both ends of the problem pipeline.} Problem-core matching can be exploited before the Ising mapping, from the problem as posed, and after it, from the delivered QUBO alone. Neither requires per-instance analysis or machinery beyond a load balancer, so \arch's advantage is reachable with simple dispatch. Extending these signals to previously unseen problem classes remains open.}

% Tmp fig. dump:
%\input{sections/evaluationBank}

\section{Related Work} \label{sec:related}
\blueHL{Section~\ref{sec:background} divides Ising solvers in hardware into those that implement search procedures directly}~\cite{raman2024sachi,yangsophie,chiang2024reaim}\blueHL{ and those that implement Ising-model-compliant physical systems. Only the second class exposes physical connectivity as a design variable, so only its members can serve as component \arch\ cores; the first remains relevant as prior work on scaling, which we discuss below.} The literature is broad and fast-moving; see~\cite{mohseni2022ising} for a survey.

\blueHL{The closest prior efforts to \arch\ scale a \emph{single} connectivity type.} Multi-chip Ising machines tile many instances of one topology to grow capacity. Tatsumura et al.~\cite{tatsumura2021scaling} scale out simulated-bifurcation machines with full spin-to-spin connectivity across chips, and Kashimata et al.~\cite{Kashimata_2024} extend this by overlapping in-chip computation with inter-chip communication to hide the resulting overhead. Others network sparse King's-graph chips~\cite{9634769} or attach a multi-chip extension to a fully connected annealer~\cite{kawamura2023amorphica}. A separate line keeps one chip but makes its connectivity reconfigurable, emulating denser coupling on sparse hardware~\cite{nikhar2024reconfig}. \blueHL{All of these hold connectivity uniform across the workload: they grow how large a problem one topology can hold, whereas \arch\ varies connectivity \emph{across} cores to match how different problems are shaped. This is also why \arch's cores do not communicate. Coupling independent cores would reintroduce the embedding and convergence penalties of a Hybrid topology at chip scale (Section~\ref{sec:trade-off}), trading away the flexibility that heterogeneity provides. A heterogeneous chip can still be replicated and scaled out without coupling its cores, so these capacity-scaling techniques compose with \arch, but none of them targets the workload-diversity question \arch\ addresses.}

\blueHL{One multi-chip design~\cite{sharma2022increasing} also supports an independent mode, in which chips solve separate smaller problems rather than cooperating on one large problem. \arch\ shares the independence but not the goal: those chips are identical, so independence buys only throughput, whereas \arch's cores differ in connectivity, and independence lets each problem run on the topology that fits it.}

Several other efforts are orthogonal to \arch\ and can be combined with it: more efficient problem decomposition~\cite{tan2023hyqsat,cilasun20243sat}; more compact problem formulation~\cite{zielinski2023pattern,nusslein2022algorithmic,Packebusch_2016,9831054,chen2024influencemaximizationisingmodels,PUERTO2022105701}; models that natively support higher-order interactions~\cite{Glaetzle2017,bashar2023designing,bybee2023efficient,sharma2023augmenting}; and hardware connectivity or precision optimization~\cite{Lo2023,puri2017quantum,kumar2020large}.

\section{Conclusion} \label{sec:conclusion}
This paper makes the case for \arch, a heterogeneous Ising multiprocessor in which independent cores of different connectivity share a chip, and a scheduler assigns each problem to the core whose connectivity fits it. \blueHL{The argument rests on three results. First, no single connectivity serves a diverse workload: the favorable topology moves with problem structure, set by the interaction graph that remains after a problem is mapped to the Ising model. Second, a middle-ground topology does not resolve this. A Hybrid core almost never attains the lowest time-to-solution, and among the configurations evaluated the strongest pair the connectivity extremes. Third, the assignment this depends on can be made from information a system actually has: we propose two schedulers, one keyed on the problem's family and parameters before the Ising mapping and one on the coupling density of the delivered QUBO after it, and both realize the benefit without per-instance analysis.}

\blueHL{To reach these results under a controlled hardware budget we built a fully digital, synthesizable Ising-machine emulator, which gives every spin and every coupling its own circuit and therefore exposes the area and routing cost of connectivity in a form that automated design flows can measure. It is the instrument that makes iso-area comparison across topologies possible without fabricating a chip per design point.}

The key contribution is an architecture-level insight that applies across Ising core designs, so any such design can serve as a component core in this context. \blueHL{The capacities and speedups reported here are those of the modeled implementation and workload set; what carries beyond them is the architectural principle rather than the numbers.} We focused on integrating cores of the same technology but diverse connectivity into a single chip; extending the same principle beyond chip boundaries, subject to hardware resource budgets, remains open.

%%%%%%% -- PAPER CONTENT ENDS -- %%%%%%%%

%%%%%%%%% -- BIB STYLE AND FILE -- %%%%%%%%
\bibliographystyle{IEEEtran}
\bibliography{refs}

@article{moy20221,
  title={A 1,968-node coupled ring oscillator circuit for combinatorial optimization problem solving},
  author={Moy, William and Ahmed, Ibrahim and Chiu, Po-wei and Moy, John and Sapatnekar, Sachin S and Kim, Chris H},
  journal={Nature Electronics},
  volume={5},
  number={5},
  pages={310--317},
  year={2022},
  publisher={Nature Publishing Group UK London}
}

@Article{Lo2023,
author={Lo, Hao
and Moy, William
and Yu, Hanzhao
and Sapatnekar, Sachin
and Kim, Chris H.},
title={An Ising solver chip based on coupled ring oscillators with a 48-node all-to-all connected array architecture},
journal={Nature Electronics},
year={2023},
month={Oct},
day={01},
volume={6},
number={10},
pages={771-778},
issn={2520-1131},
doi={10.1038/s41928-023-01021-y},
url={https://doi.org/10.1038/s41928-023-01021-y}
}

@article{ebadi2022quantum,
  title={Quantum optimization of maximum independent set using Rydberg atom arrays},
  author={Ebadi, Sepehr and Keesling, Alexander and Cain, Madelyn and Wang, Tout T and Levine, Harry and Bluvstein, Dolev and Semeghini, Giulia and Omran, Ahmed and Liu, J-G and Samajdar, Rhine and others},
  journal={Science},
  volume={376},
  number={6598},
  pages={1209--1215},
  year={2022},
  publisher={American Association for the Advancement of Science}
}

@article{dutta2021ising,
  title={An Ising Hamiltonian solver based on coupled stochastic phase-transition nano-oscillators},
  author={Dutta, Suryendy and Khanna, Abhishek and Assoa, AS and Paik, Hanjong and Schlom, Darrell G and Toroczkai, Zolt{\'a}n and Raychowdhury, Arijit and Datta, Suman},
  journal={Nature Electronics},
  volume={4},
  number={7},
  pages={502--512},
  year={2021},
  publisher={Nature Publishing Group UK London}
}

@misc{takemoto20192,
  title={2.6 A 2$\times$ 30k-spin multichip scalable annealing processor based on a processing-in-memory approach for solving large-scale combinatorial optimization problems},
  author={Takemoto, Takashi and Hayashi, Masato and Yoshimura, Chihiro and Yamaoka, Masanao},
  booktitle={Digest of Technical Papers, IEEE International Solid-State Circuits Conference (ISSCC)},
  pages={52--54},
  year={2019},
  organization={IEEE}
}

@article{inagaki2016coherent,
  title={A coherent {Ising} machine for 2000-node optimization problems},
  author={Inagaki, Takahiro and Haribara, Yoshitaka and Igarashi, Koji and Sonobe, Tomohiro and Tamate, Shuhei and Honjo, Toshimori and Marandi, Alireza and McMahon, Peter L and Umeki, Takeshi and Enbutsu, Koji and others},
  journal={Science},
  volume={354},
  number={6312},
  pages={603--606},
  year={2016},
  publisher={American Association for the Advancement of Science}
}

@article{yamamoto2020statica,
  title={STATICA: A 512-spin 0.25 M-weight annealing processor with an all-spin-updates-at-once architecture for combinatorial optimization with complete spin--spin interactions},
  author={Yamamoto, Kasho and Kawamura, Kazushi and Ando, Kota and Mertig, Normann and Takemoto, Takashi and Yamaoka, Masanao and Teramoto, Hiroshi and Sakai, Akira and Takamaeda-Yamazaki, Shinya and Motomura, Masato},
  journal={IEEE Journal of Solid-State Circuits},
  volume={56},
  number={1},
  pages={165--178},
  year={2020},
  publisher={IEEE}
}

@article{goto2021high,
  title={High-performance combinatorial optimization based on classical mechanics},
  author={Goto, Hayato and Endo, Kotaro and Suzuki, Masaru and Sakai, Yoshisato and Kanao, Taro and Hamakawa, Yohei and Hidaka, Ryo and Yamasaki, Masaya and Tatsumura, Kosuke},
  journal={Science Advances},
  volume={7},
  number={6},
  pages={eabe7953},
  year={2021},
  publisher={American Association for the Advancement of Science}
}

@article{tatsumura2021scaling,
  title={Scaling out Ising machines using a multi-chip architecture for simulated bifurcation},
  author={Tatsumura, Kosuke and Yamasaki, Masaya and Goto, Hayato},
  journal={Nature Electronics},
  volume={4},
  number={3},
  pages={208--217},
  year={2021},
  publisher={Nature Publishing Group UK London}
}

@article{aramon2019physics,
  title={Physics-inspired optimization for quadratic unconstrained problems using a digital annealer},
  author={Aramon, Maliheh and Rosenberg, Gili and Valiante, Elisabetta and Miyazawa, Toshiyuki and Tamura, Hirotaka and Katzgraber, Helmut G},
  journal={Frontiers in Physics},
  volume={7},
  pages={48},
  year={2019},
  publisher={Frontiers Media SA}
}

@article{nakayama2021description,
  title={Description: third generation digital annealer technology},
  author={Nakayama, Hiroshi and Koyama, Junpei and Yoneoka, Noboru and Miyazawa, Toshiyuki},
  journal={Fujitsu Limited: Tokyo, Japan},
  year={2021}
}

@article{clements2017gaussian,
  title={Gaussian optical Ising machines},
  author={Clements, William R and Renema, Jelmer J and Wen, Y Henry and Chrzanowski, Helen M and Kolthammer, W Steven and Walmsley, Ian A},
  journal={Physical Review A},
  volume={96},
  number={4},
  pages={043850},
  year={2017},
  publisher={APS}
}

@article{pierangeli2019large,
  title={Large-scale photonic Ising machine by spatial light modulation},
  author={Pierangeli, D and Marcucci, G and Conti, C},
  journal={Physical Review Letters},
  volume={122},
  number={21},
  pages={213902},
  year={2019},
  publisher={APS}
}

@ARTICLE{isingNPHard,
       author = {{Barahona}, F.},
        title = "{On the computational complexity of Ising spin glass models}",
      journal = {Journal of Physics A: Mathematical and General},
         year = 1982,
        month = oct,
       volume = {15},
       number = {10},
        optpages = {3241-3253},
          optdoi = {10.1088/0305-4470/15/10/028},
}

@article{sa,
	author = {Kirkpatrick, S. and Gelatt, C. D. and Vecchi, M. P.},
	title = {Optimization by Simulated Annealing},
	volume = {220},
	number = {4598},
	pages = {671--680},
	year = {1983},
	optdoi = {10.1126/science.220.4598.671},
	journal = {Science}
}

@ARTICLE{lucas14,
 AUTHOR={Lucas, Andrew},   
 TITLE={Ising formulations of many NP problems},      
 JOURNAL={Frontiers in Physics},      
 VOLUME={2},      
 OPTPAGES={5},     
 YEAR={2014},      
}

@misc{isingBasicsDWave,
  title={The {Ising} model: teaching an old problem new tricks},
  author={Zhengbing Bian and Fabi{\'a}n A. Chudak and William G. Macready and Geordie Rose},
  note= {DWave Systems},
  year={2010}
}

@article{dwave1,
title = "Quantum annealing with manufactured spins",
author = "Johnson, {M. W.} and Amin, {M. H S} and S. Gildert and T. Lanting and F. Hamze and N. Dickson and R. Harris and Berkley, {A. J.} and J. Johansson and P. Bunyk and Chapple, {E. M.} and C. Enderud and Hilton, {J. P.} and K. Karimi and E. Ladizinsky and N. Ladizinsky and T. Oh and I. Perminov and C. Rich and Thom, {M. C.} and E. Tolkacheva and Truncik, {C. J S} and S. Uchaikin and J. Wang and B. Wilson and G. Rose",
year = "2011",
month = may,
doi = "10.1038/nature10012",
volume = "473",
optpages = "194--198",
journal = "Nature",
number = "7346",
}

@book{i, author = {Cook, William J. and Cunningham, William H. and Pulleyblank, William R. and Schrijver, Alexander}, title = {Combinatorial Optimization}, year = {1998}, isbn = {047155894X}, publisher = {John Wiley \& Sons, Inc.}, address = {USA} }

@article{v,
	month = {jun},
	publisher = {{IOP} Publishing},
	volume = {19},
	number = {9},
	optpages = {1605--1620},
	author = {Y Fu and P W Anderson},
	title = {Application of statistical mechanics to {NP}-complete problems in combinatorial optimisation},
	journal = {Journal of Physics A: Mathematical and General},
    year = 1986
}

@article{vi,
author = {Barahona, Francisco and Gr\"{o}tschel, Martin and J\"{u}nger, Michael and Reinelt, Gerhard},
title = {An Application of Combinatorial Optimization to Statistical Physics and Circuit Layout Design},
year = {1988},
issue_date = {June 1988},
publisher = {INFORMS},
volume = {36},
number = {3},
journal = {Oper. Res.},
month = jun,
optpages = {493–513},
numpages = {21},
}

@article {viii,
	author = {Farhi, Edward and Goldstone, Jeffrey and Gutmann, Sam and Lapan, Joshua and Lundgren, Andrew and Preda, Daniel},
	title = {A Quantum Adiabatic Evolution Algorithm Applied to Random Instances of an NP-Complete Problem},
	volume = {292},
	number = {5516},
	optpages = {472--475},
	year = {2001},
	journal = {Science}
}

@article{palubeckis2004multistart,
  title={Multistart tabu search strategies for the unconstrained binary quadratic optimization problem},
  author={Palubeckis, Gintaras},
  journal={Annals of Operations Research},
  volume={131},
  pages={259--282},
  year={2004},
  publisher={Springer}
}

@article{chancellor2016direct,
  title={A direct mapping of Max k-SAT and high order parity checks to a chimera graph},
  author={Chancellor, Nicholas and Zohren, Stefan and Warburton, Paul A and Benjamin, Simon C and Roberts, Stephen},
  journal={Scientific Reports},
  volume={6},
  number={1},
  pages={37107},
  year={2016},
  publisher={Nature Publishing Group UK London}
}

@inproceedings{bothers,
  title={What’s Wrong with Deep Learning in Tree Search for Combinatorial Optimization},
  author={B{\"o}ther, Maximilian and Ki{\ss}ig, Otto and Taraz, Martin and Cohen, Sarel and Seidel, Karen and Friedrich, Tobias},
  booktitle={International Conference on Learning Representations}, 
year={2022}
}

@article{holme2002growing,
  title={Growing scale-free networks with tunable clustering},
  author={Holme, Petter and Kim, Beom Jun},
  journal={Physical review E},
  volume={65},
  number={2},
  pages={026107},
  year={2002},
  publisher={APS}
}

@inproceedings{sharma2022increasing,
  title={Increasing ising machine capacity with multi-chip architectures},
  author={Sharma, Anshujit and Afoakwa, Richard and Ignjatovic, Zeljko and Huang, Michael},
  booktitle={Proceedings of the 49th Annual International Symposium on Computer Architecture},
  pages={508--521},
  year={2022}
}

@article{cilasun20243sat,
  title={3SAT on an all-to-all-connected CMOS Ising solver chip},
  author={C{\i}lasun, H{\"u}srev and Zeng, Ziqing and Kumar, Abhimanyu and Lo, Hao and Cho, William and Moy, William and Kim, Chris H and Karpuzcu, Ulya R and Sapatnekar, Sachin S},
  journal={Scientific reports},
  volume={14},
  number={1},
  pages={10757},
  year={2024},
  publisher={Nature Publishing Group UK London}
}

@article{Boixo_2014,
   title={Evidence for quantum annealing with more than one hundred qubits},
   volume={10},
   ISSN={1745-2481},
   url={http://dx.doi.org/10.1038/nphys2900},
   DOI={10.1038/nphys2900},
   number={3},
   journal={Nature Physics},
   publisher={Springer Science and Business Media LLC},
   author={Boixo, Sergio and Rønnow, Troels F. and Isakov, Sergei V. and Wang, Zhihui and Wecker, David and Lidar, Daniel A. and Martinis, John M. and Troyer, Matthias},
   year={2014},
   month=feb, pages={218–224} }

@INPROCEEDINGS{9634769,
  author={Yamamoto, Kasho and Takemoto, Takashi and Yoshimura, Chihiro and Mashimo, Mayumi and Yamaoka, Masanao},
  booktitle={2021 IEEE Asian Solid-State Circuits Conference (A-SSCC)}, 
  title={A 1.3-Mbit Annealing System Composed of Fully-Synchronized 9-board x 9-chip x 16-kbit Annealing Processor Chips for Large-Scale Combinatorial Optimization Problems}, 
  year={2021},
  volume={},
  number={},
  pages={1-3},
  doi={10.1109/A-SSCC53895.2021.9634769}}

@article{Kashimata_2024,
   title={Efficient and Scalable Architecture for Multiple-Chip Implementation of Simulated Bifurcation Machines},
   volume={12},
   ISSN={2169-3536},
   url={http://dx.doi.org/10.1109/ACCESS.2024.3374089},
   DOI={10.1109/access.2024.3374089},
   journal={IEEE Access},
   publisher={Institute of Electrical and Electronics Engineers (IEEE)},
   author={Kashimata, Tomoya and Yamasaki, Masaya and Hidaka, Ryo and Tatsumura, Kosuke},
   year={2024},
   pages={36606–36621} }

@article{Packebusch_2016,
   title={Low autocorrelation binary sequences},
   volume={49},
   ISSN={1751-8121},
   url={http://dx.doi.org/10.1088/1751-8113/49/16/165001},
   DOI={10.1088/1751-8113/49/16/165001},
   number={16},
   journal={Journal of Physics A: Mathematical and Theoretical},
   publisher={IOP Publishing},
   author={Packebusch, Tom and Mertens, Stephan},
   year={2016},
   month=mar, pages={165001} }

@ARTICLE{9831054,
  author={Singh, Abhishek Kumar and Jamieson, Kyle and McMahon, Peter L. and Venturelli, Davide},
  journal={IEEE Transactions on Wireless Communications}, 
  title={Ising Machines’ Dynamics and Regularization for Near-Optimal MIMO Detection}, 
  year={2022},
  volume={21},
  number={12},
  pages={11080-11094},
  doi={10.1109/TWC.2022.3189604}}

@misc{chen2024influencemaximizationisingmodels,
      title={Influence Maximization in Ising Models}, 
      author={Zongchen Chen and Elchanan Mossel},
      year={2024},
      eprint={2309.05206},
      archivePrefix={arXiv},
      primaryClass={cs.DS},
      url={https://arxiv.org/abs/2309.05206}, 
}

@article{PUERTO2022105701,
  title={A combinatorial optimization approach to scenario filtering in portfolio selection},
  author={Puerto, Justo and Ricca, Federica and Rodr{\'\i}guez-Madrena, Mois{\'e}s and Scozzari, Andrea},
  journal={Computers \& Operations Research},
  volume={142},
  pages={105701},
  year={2022},
  publisher={Elsevier}
}

@article{Glaetzle2017,
  author = {Glaetzle, Alexander W. and van Bijnen, Rick M. W. and Zoller, Peter and Lechner, Wolfgang and Buchler, Hans Peter and Dalmonte, Marcello and Pupillo, Guido},
  title = {A coherent quantum annealer with Rydberg atoms},
  journal = {Nature Communications},
  volume = {8},
  pages = {15813},
  year = {2017},
  doi = {10.1038/ncomms15813},
  url = {https://doi.org/10.1038/ncomms15813}
}

@article{sharma2023augmenting,
  title={Augmenting an electronic Ising machine to effectively solve boolean satisfiability},
  author={Sharma, Anshujit and Burns, Matthew and Hahn, Andrew and Huang, Michael},
  journal={Scientific Reports},
  volume={13},
  number={1},
  pages={22858},
  year={2023},
  publisher={Nature Publishing Group UK London}
}

@article{bybee2023efficient,
  title={Efficient optimization with higher-order Ising machines},
  author={Bybee, Connor and Kleyko, Denis and Nikonov, Dmitri E and Khosrowshahi, Amir and Olshausen, Bruno A and Sommer, Friedrich T},
  journal={Nature Communications},
  volume={14},
  number={1},
  pages={6033},
  year={2023},
  publisher={Nature Publishing Group UK London}
}

@article{bashar2023designing,
  title={Designing Ising machines with higher order spin interactions and their application in solving combinatorial optimization},
  author={Bashar, Mohammad Khairul and Shukla, Nikhil},
  journal={Scientific Reports},
  volume={13},
  number={1},
  pages={9558},
  year={2023},
  publisher={Nature Publishing Group UK London}
}

@inproceedings{tan2023hyqsat,
  title={HyQSAT: A hybrid approach for 3-SAT problems by integrating quantum annealer with CDCL},
  author={Tan, Siwei and Yu, Mingqian and Python, Andre and Shang, Yongheng and Li, Tingting and Lu, Liqiang and Yin, Jianwei},
  booktitle={2023 IEEE International Symposium on High-Performance Computer Architecture (HPCA)},
  pages={731--744},
  year={2023},
  organization={IEEE}
}

@article{zielinski2023pattern,
  title={Pattern QUBOs: Algorithmic construction of 3SAT-to-QUBO transformations},
  author={Zielinski, Sebastian and N{\"u}{\ss}lein, Jonas and Stein, Jonas and Gabor, Thomas and Linnhoff-Popien, Claudia and Feld, Sebastian},
  journal={Electronics},
  volume={12},
  number={16},
  pages={3492},
  year={2023},
  publisher={MDPI}
}

@inproceedings{nusslein2022algorithmic,
  title={Algorithmic QUBO formulations for k-SAT and hamiltonian cycles},
  author={N{\"u}{\ss}lein, Jonas and Gabor, Thomas and Linnhoff-Popien, Claudia and Feld, Sebastian},
  booktitle={Proceedings of the genetic and evolutionary computation conference companion},
  pages={2240--2246},
  year={2022}
}

@article{puri2017quantum,
  title={Quantum annealing with all-to-all connected nonlinear oscillators},
  author={Puri, Shruti and Andersen, Christian Kraglund and Grimsmo, Arne L and Blais, Alexandre},
  journal={Nature communications},
  volume={8},
  number={1},
  pages={15785},
  year={2017},
  publisher={Nature Publishing Group UK London}
}

@article{kumar2020large,
  title={Large-scale Ising emulation with four body interaction and all-to-all connections},
  author={Kumar, Santosh and Zhang, He and Huang, Yu-Ping},
  journal={Communications Physics},
  volume={3},
  number={1},
  pages={108},
  year={2020},
  publisher={Nature Publishing Group UK London}
}

@inproceedings{raman2024sachi,
  title={SACHI: A Stationarity-Aware, All-Digital, Near-Memory, Ising Architecture},
  author={Raman, Siddhartha Raman Sundara and John, Lizy K and Kulkarni, Jaydeep P},
  booktitle={2024 IEEE International Symposium on High-Performance Computer Architecture (HPCA)},
  pages={719--731},
  year={2024},
  organization={IEEE}
}

@inproceedings{yangsophie,
  title={{SOPHIE}: A Scalable Recurrent Ising Machine Using Optically Addressed Phase Change Memory},
  author={Yang, Guowei and Karimi, Sina and Ocampo, Carlos A R{\'\i}os and Coskun, Ayse K and Joshi, Ajay},
  booktitle={2024 57th IEEE/ACM International Symposium on Microarchitecture (MICRO)},
  year={2024},
  organization={IEEE}
}

@inproceedings{chiang2024reaim,
  title={ReAIM: A ReRAM-based Adaptive Ising Machine for Solving Combinatorial Optimization Problems},
  author={Chiang, Hao-Wei and Nien, Chin-Fu and Cheng, Hsiang-Yun and Huang, Kuei-Po},
  booktitle={2024 ACM/IEEE 51st Annual International Symposium on Computer Architecture (ISCA)},
  pages={58--72},
  year={2024},
  organization={IEEE}
}

@inproceedings{mallick2021overcoming,
  title={Overcoming the accuracy vs. performance trade-off in oscillator ising machines},
  author={Mallick, A and Bashar, MK and Truesdell, DS and Calhoun, BH and Shukla, N},
  booktitle={2021 IEEE International Electron Devices Meeting (IEDM)},
  pages={40--2},
  year={2021},
  organization={IEEE}
}

@article{erdos1959random,
  title        = {On {R}andom {G}raphs {I}},
  author       = {Erd{\H{o}}s, Paul and R{\'e}nyi, Alfr{\'e}d},
  journal      = {Publicationes Mathematicae},
  volume       = {6},
  pages        = {290--297},
  year         = {1959}
}

@article{newman2003structure,
  title        = {The Structure and Function of Complex Networks},
  author       = {Newman, M. E. J.},
  journal      = {SIAM Review},
  volume       = {45},
  number       = {2},
  pages        = {167--256},
  year         = {2003},
  publisher    = {Society for Industrial and Applied Mathematics}
}

@inproceedings{eikmeier2017revisiting,
  title        = {Revisiting Power-law Distributions in Spectra of Real World Networks},
  author       = {Eikmeier, Nicole and Gleich, David F.},
  booktitle    = {Proceedings of the 23rd {ACM} {SIGKDD} International Conference on Knowledge Discovery and Data Mining},
  pages        = {817--826},
  year         = {2017},
  publisher    = {ACM}
}

@article{barabasi1999emergence,
  title        = {Emergence of Scaling in Random Networks},
  author       = {Barab{\'a}si, Albert-L{\'a}szl{\'o} and Albert, R{\'e}ka},
  journal      = {Science},
  volume       = {286},
  number       = {5439},
  pages        = {509--512},
  year         = {1999},
  publisher    = {American Association for the Advancement of Science}
}

@article{cilasun2025cobi,
  author    = {H{\"u}srev C{\i}lasun and William Moy and Ziqing Zeng and Tahmida Islam and Hao Lo and Alex Vanasse and Megan Tan and Mohammad Anees and Ramprasath S and Abhimanyu Kumar and Sachin S. Sapatnekar and Chris H. Kim and Ulya R. Karpuzcu},
  title     = {A coupled-oscillator-based {Ising} chip for combinatorial optimization},
  journal   = {Nature Electronics},
  volume    = {8},
  number    = {6},
  pages     = {537--546},
  year      = {2025},
  doi       = {10.1038/s41928-025-01393-3},
}

@article{kumar2025droid,
  author    = {Abhimanyu Kumar and Ramprasath S. and Chris H. Kim and Ulya R. Karpuzcu and Sachin S. Sapatnekar},
  title     = {{DROID}: discrete-time simulation for ring-oscillator-based {Ising} design},
  journal   = {Scientific Reports},
  volume    = {15},
  pages     = {18643},
  year      = {2025},
  doi       = {10.1038/s41598-025-00037-y},
}

@inproceedings{su2017fast,
  author    = {J. Su and L. He},
  title     = {Fast embedding of constrained satisfaction problem to quantum annealer with minimizing chain length},
  booktitle = {54th ACM/EDAC/IEEE Design Automation Conference (DAC)},
  year      = {2017},
}

@inproceedings{su2016quantum,
  author    = {J. Su and T. Tu and L. He},
  title     = {A quantum annealing approach for {Boolean} Satisfiability problem},
  booktitle = {53rd ACM/EDAC/IEEE Design Automation Conference (DAC)},
  year      = {2016},
}

@article{cai2014practical,
  author    = {Jun Cai and William G. Macready and Aidan Roy},
  title     = {A practical heuristic for finding graph minors},
  journal   = {arXiv preprint arXiv:1406.2741},
  year      = {2014},
}

@article{hamerly2019experimental,
  author    = {Ryan Hamerly and Takahiro Inagaki and Peter L. McMahon and Davide Venturelli and Alireza Marandi and Tatsuhiro Onodera and Edwin Ng and Carsten Langrock and Kensuke Inaba and Toshimori Honjo and Koji Enbutsu and Takeshi Umeki and Ryoichi Kasahara and Shoko Utsunomiya and Satoshi Kako and Ken-ichi Kawarabayashi and Robert L. Byer and Martin M. Fejer and Hideo Mabuchi and Dirk Englund and Eleanor Rieffel and Hiroki Takesue and Yoshihisa Yamamoto},
  title     = {Experimental investigation of performance differences between coherent {Ising} machines and a quantum annealer},
  journal   = {Science Advances},
  volume    = {5},
  number    = {5},
  pages     = {eaau0823},
  year      = {2019},
  doi       = {10.1126/sciadv.aau0823},
}

@article{konz2021embedding,
  author    = {Mario S. K{\"o}nz and Wolfgang Lechner and Helmut G. Katzgraber and Matthias Troyer},
  title     = {Embedding Overhead Scaling of Optimization Problems in Quantum Annealing},
  journal   = {PRX Quantum},
  volume    = {2},
  pages     = {040322},
  year      = {2021},
  doi       = {10.1103/PRXQuantum.2.040322},
}

@misc{booth2017partitioning,
  author    = {Michael Booth and Steven P. Reinhardt and Aidan Roy},
  title     = {Partitioning Optimization Problems for Hybrid Classical/Quantum Execution},
  year      = {2017},
  note      = {D-Wave Technical Report 14-1006A-A},
}

@inproceedings{shalan2020openlane,
  author    = {Mohamed Shalan and Tim Edwards},
  title     = {Building {OpenLANE}: A 130nm {OpenROAD}-based 
               Tapeout-Proven Flow},
  booktitle = {IEEE/ACM International Conference on Computer-Aided 
               Design (ICCAD)},
  pages     = {1--6},
  year      = {2020},
}

@article{ajayi2019openroad,
  author    = {Tutu Ajayi and Vidya A. Chhabria and Mateus Fogaca 
               and Soheil Hashemi and Abdelrahman Hosny and Andrew 
               B. Kahng and Minsoo Kim and Jeongsup Lee and Uday 
               Mallappa and Marina Neseem and Geraldo Pradipta and 
               Sherief Reda and Mehdi Saligane and Sachin S. 
               Sapatnekar and Carl Sechen and Mohamed Shalan and 
               William Swartz and Lutong Wang and Zhehui Wang and 
               Mingyu Woo and Bangqi Xu},
  title     = {{OpenROAD}: Toward a Self-Driving, Open-Source 
               Digital Layout Implementation Tool Chain},
  journal   = {Government Microcircuit Applications and Critical 
               Technology Conference (GOMACTech)},
  year      = {2019},
}

@article{wang2017oim,
  title={Oscillator-based Ising Machine},
  author={Wang, Tianshi and Roychowdhury, Jaijeet},
  journal={arXiv preprint arXiv:1709.08102},
  year={2017}
}

@article{nikhar2024reconfig,
  title={All-to-all reconfigurability with sparse and higher-order Ising machines},
  author={Nikhar, Srijan and Kannan, Shuvro and Aadit, Navid Anjum and Chowdhury, Shuvro and Camsari, Kerem Y.},
  journal={Nature Communications},
  volume={15},
  number={1},
  pages={8977},
  year={2024},
  publisher={Nature Publishing Group},
  doi={10.1038/s41467-024-53270-w}
}

@article{mohseni2022ising,
  title={Ising machines as hardware solvers of combinatorial optimization problems},
  author={Mohseni, Naeimeh and McMahon, Peter L. and Byrnes, Tim},
  journal={Nature Reviews Physics},
  volume={4},
  number={6},
  pages={363--379},
  year={2022},
  publisher={Nature Publishing Group},
  doi={10.1038/s42254-022-00440-8}
}

@inproceedings{kawamura2023amorphica,
  title={Amorphica: 4-Replica 512 Fully Connected Spin 336MHz Metamorphic Annealer with Programmable Optimization Strategy and Compressed-Spin-Transfer Multi-Chip Extension},
  author={Kawamura, Kazushi and Yu, Jaehoon and Okonogi, Daiki and Jimbo, Satoru and Inoue, Genta and Hyodo, Akira and L{\'o}pez Garc{\'i}a-Arias, {\'A}ngel and Ando, Kota and Fukushima-Kimura, Bruno Hideki and Yasudo, Ryota and Van Chu, Thiem and Motomura, Masato},
  booktitle={2023 IEEE International Solid-State Circuits Conference (ISSCC)},
  pages={42--44},
  year={2023},
  organization={IEEE}
}
%%%%%%%%%%%%%%%%%%%%%%%%%%%%%%%%%%%%

\end{document}